\documentclass[5p,number,times]{elsarticle}

\usepackage[intlimits,sumlimits]{amsmath}
\usepackage[nottoc]{tocbibind}
\usepackage{amsfonts}
\usepackage{siunitx}
\usepackage{graphicx}
\usepackage[english]{babel}
\usepackage{isomath}
\usepackage{color}
\usepackage{hyperref}
\hypersetup{breaklinks=true,colorlinks=true,citecolor=blue,pdfstartview={FitH}, linkcolor=blue}
\usepackage[mathscr]{euscript}
\usepackage{mathtools}
\usepackage{float}
\biboptions{authoryear}%chicago}%elsarticle-harv}
\def\({\left(}
\def\){\right)}
\def\[{\left[ }
\def\]{\right]}

\def\d{ \mathrm{d}}

\def\dg{$^\circ$}

\def\1{\mathbb{1}}

\def\~{\ensuremath{\sim}}
\newcommand{\ud}{\mathrm{d}}

\newcommand{\dd}[3][]{{\frac{\d^{#1} #2}{\d #3^{#1}}}}

\newcommand{\pp}[3][]{{\frac{\partial^{#1} #2}{\partial #3^{#1}}}}
\newcommand{\ppf}[3][]{{\partial^{#1} #2}/{\partial #3^{#1}}}

\newcommand{\ve}[1]{\mathrm{\mathbf{#1}}} %\usepackage{bm}  this might work better than the above

\begin{document}

\begin{abstract}
Simulating oil transport in the ocean can be done successfully provided that accurate ocean currents and surface winds are available---this is often too big of a challenge. Deficient ocean currents can sometimes be remediated by parameterizing missing physics---this is often not enough. In this chapter, we focus on some of the main problems oil-spill modelers face, which is determining accurate trajectories when the velocity may be missing important physics, or when the velocity has localized errors that result in large trajectory errors. A foundation of physical mechanisms driving motion in the ocean may help identify currents lacking certain types of physics, and the remedy. Recent progress in our understanding of motion in the upper centimeters of the ocean supports unconventional parameterizations; we present as an example the 2003 Point Wells oil spill which had remained unexplained until recently.   When the velocity realistically represents trajectory forcing mechanisms, advanced Lagrangian techniques that build on the theory of Lagrangian Coherent Structures can bypass localized velocity errors by identifying regions of attraction likely to dictate fluid deformation.  The usefulness of Objective Eulerian Coherent Structures is demonstrated to the oil-spill modeling community by revisiting the 2010 Deepwater Horizon accident in the Gulf of Mexico and predicting a prominent transport pattern from an imperfect altimetry velocity eight days in advance.
\end{abstract}

\begin{frontmatter}
%\titlepage
\title{Horizontal transport in oil-spill modeling\\ }

\author[cu1,cu2]{Rodrigo Duran\corref{c1}}
\ead{r.duran@theissresearch.org}
\cortext[c1]{Corresponding author.}

%\cortext[c2]{}
\author[c3,c3b]{Tor Nordam}
\author[c4,c4b]{Mattia Serra}
\author[c5]{Christopher Barker}

\address[cu1]{National Energy Technology Laboratory, U.S. Department of Energy, Albany OR, USA}
\address[cu2]{Theiss Research, La Jolla CA, USA}
\address[c3]{SINTEF Ocean, Trondheim, Norway}
\address[c3b]{Department of Physics, Norwegian University of Science and Technology, Trondheim, Norway}
\address[c4]{School of Engineering and Applied Sciences, Harvard University, Cambridge MA, USA}
\address[c4b]{Department of Physics, University of California San Diego, La Jolla CA, USA}

\address[c5]{Office of Response and Restoration, Emergency Response Division, National Oceanic and Atmospheric Administration, Seattle WA, USA}

\end{frontmatter}

\pagebreak

\tableofcontents{}

\pagebreak

\section{Introduction} %reviewed with grammarly, low engagement otherwise good
Successfully forecasting the movement of oil during an oil spill, along with knowing oil's current location, are the key ingredients needed to solving one of the most decisive problems during emergency response operations: where is the oil heading?  Oil's current and future locations are critical for emergency planning and response, including recovery and containment. Reconstructing past spills through hindcasts is also needed for modeling improvements, environmental impact assessments, and forensics.  The challenge of successfully forecasting or hindcasting an oil spill is a considerable one. There is no guarantee that a simulation will be successful, and deviations between simulated transport and observations are the norm rather than the exception. This chapter describes some of the main reasons why accurately simulating trajectories in the ocean is complicated and presents recent progress along different fronts that help remediate some of the problems. The focus is on hindcasts and forecasts, characterized by the need to replicate or anticipate observed oil transport. 

One of the problems is related to local deficiencies in the velocity that propagate during integration when computing trajectories, these errors often grow exponentially due to the unstable nature of ocean circulation. For this problem we present modern techniques called Objective Eulerian Coherent Structures \citep[OECS;][]{Serra2016} that can bypass errors in the velocity, identifying instantaneous attracting regions that influence transport exceptionally, and that are computed without the need to integrate the velocity. This method allows trajectory forecasting without future information, but can also be applied to ocean-model forecasts to limit the effect of discrepancies between the forecasted flow and reality.

The other type of problem we examine is related to a velocity assumed to simulate oil's motion, but that is missing some of the physical processes driving observed motion. The focus is in the upper centimeters of the ocean where atmospheric influence is strongest. The solution is to parametrize physics that drive motion when necessary. Studies in the last few years have improved our understanding of the velocity within a fine surface layer, providing information that is somewhat at odds with common practice in oil-spill modeling. As an example, we show how progress in our understanding can explain the transport of oil during the 2003 Point Wells spill in the Salish Sea, a spill that had remained unexplained for 15 years.  Also discussed are recent studies showing that parametrized near-surface processes are often responsible for oil beaching, and finally, the potential for surface convergences driven by subduction at the sea surface to indicate regions of interest for oil recovery is suggested.

A brief review of the basics of oil-spill modeling is given in section \ref{sec:two}, section \ref{sec:three} overviews transport in the ocean, the unstable nature of ocean currents, and a description of velocity products typically available to simulate oil transport. Section \ref{sec:four} is about motion in the upper layer of the ocean, describing recent progress due to numerical simulations and unique at-sea experiments.    Section \ref{sec:seven} describes some of the novel tools that may help overcome velocity errors that would result in erroneous trajectories, and exemplify their use by revisiting transport patterns observed during the Deepwater Horizon accident in May 2010. We conclude in section \ref{sec:eight} with conclusions and an outlook of what progress can be expected, and how the techniques presented here fit into that picture.

%; some comments are given for the different, related problem of running an ensemble of oil-spill simulations to assess environmental impacts statistically.

\section{The physics, the mathematics and the numerics} %reviewed with grammarly very clear just right, engaging overall 85
\label{sec:two}
Oil-spill modeling is often a multidisciplinary endeavor; it is not uncommon for biologists, chemists, physicists, geographers, mathematicians, oceanographers, engineers, and computer scientists to work together.  It is therefore helpful to begin clarifying, somewhat informally, the basic physics, mathematics, and numerical solutions used to simulate the transport in a fluid that ultimately results in an oil-spill simulation.  The physical mechanisms that drive motion in the ocean are described in later sections.

The physical approximations used to simulate transport forced by a vast variety of oceanographic processes are well known. Consequently, the mathematical equations are also well known (e.g. studied in most introductory partial differential equation courses). The mathematics of oil transport boils down to solving the advection-diffusion equation, also known as the dispersion-diffusion equation.  In fluid dynamics and physical oceanography, the effects of advection and diffusion are often referred to as stirring and mixing, respectively. In oil-spill modeling, these equations are solved in a Lagrangian framework, i.e. by computing the trajectories of individual elements ("particles"). However, it is illustrative to introduce the equations of mathematical physics in Eulerian terms, i.e. as a function of fixed space, and return to the Lagrangian formulation when discussing the numerical solution.

Stirring within a fluid causes a tracer to deform into streaks and swirls while the along-path concentration of the tracer does not change. Effectively, the tracer is redistributed with the velocity field while preserving its concentration, such behavior is called conservative. Assuming the velocity divergence is negligible,  the following evolution equation for a tracer $C$ is satisfied:
\begin{equation}
\pp{C}{t} + \ve{u} \cdot \nabla C = 0
\label{eq:adv}
\end{equation}
where $\ve{u}$ is the two-dimensional velocity of the fluid. This equation is known as the advection equation or dispersion equation. Some scientists may use the word convection instead of advection, although oceanographers often reserve the term convection for a different type of physics (vertical motion related to buoyancy changes).  The advection equation describes how a concentration evolves as the velocity field acts upon the gradient of the concentration, causing a redistribution of the tracer. \\

%When working with a two-dimensional velocity, there may a loss (sink) of mass due to a vertical flux.  As we will see in section \ref{sec:fronts}, this is not uncommon at the sea surface. In this case, changes in concentration due to the velocity divergence can be simulated with the full advection equation:
%
%\begin{equation}
%\pp{C}{t} +   \nabla \cdot \left( C\ve{u} \right) = 0
%\label{eq:advfull}
%\end{equation}

Mixing refers to the diffusion of a tracer with concentration $C$, typically simulated with the diffusion equation:
\begin{equation}
     \pp{C}{t} = \nabla \cdot \left( \kappa \nabla C \right)
\label{eq:diff}
\end{equation}

If $\kappa$ is a molecular diffusion coefficient, then \eqref{eq:diff} represents the mixing of a tracer due to molecular collisions. By itself, this is an inefficient method of mixing a fluid. In the ocean, however, an eddy diffusion coefficient is used for $\kappa$, several orders of magnitude larger than the molecular coefficient that is characteristic of the fluid. From a physics point of view, the large coefficient means that the diffusion equation is modeling the mixing of concentration due to the collision of water parcels, not molecules. This is an ad hoc way of parameterizing small-scale stirring and overturning of water parcels as they undergo turbulent motions. Ironically,  the physics simulated by equation \eqref{eq:diff}  are well understood, yet the ocean physics that it parametrizes includes a variety of processes that are difficult to even measure \cite[e.g.][]{Moum2009}.  It is a fortunate turn of events that results from using a well-understood equation such as \eqref{eq:diff}  are adequate for many purposes, including oil-spill modeling.

To simulate transport in the ocean, both stirring and diffusion are often used simultaneously: the evolution equation for the transport of a tracer is then the advection-diffusion (or dispersion-diffusion) equation:
\begin{equation}
\pp{C}{t} + \ve{u} \cdot \nabla C = \nabla \cdot \left( \kappa \nabla C \right)
\label{eq:adv-diff}
\end{equation}

In the case of oil-spill modeling, an additional source term can be added to represent the addition of oil as it is spilling into the ocean; a sink term can also be included to represent the removal of oil. This chapter is concerned with the transport of oil, and we will not consider sources or sinks. \cite{Salmon1998} presents a discussion (his section 14) on stirring and mixing, describing their individual and simultaneous effects on tracer variability. Stirring, diffusion, and their interplay is also discussed in section 7.3 of \cite{Tennekes1972}.

The eddy diffusion coefficient, also known as the turbulent diffusion coefficient, is an unknown that needs to be determined. The production of turbulence, and therefore the magnitude of the eddy diffusion coefficient, depends on many factors including the spatial structure of seawater's density, the spatial structure of the velocity field, heating or cooling of water parcels, and Earth's rotation. Turbulence in the ocean is often produced by instabilities that range in length from centimeters to hundreds of kilometers---the interested reader is referred to the free book on ocean instabilities by \cite{smyth2019}.  Due to a large number of instability types, the large range of spatial and temporal scales at which they occur and interact, and their often anisotropic nature, determining an adequate turbulent diffusion coefficient is a difficult problem.

Numerical ocean models require accurate mixing of momentum, heat, and salinity to produce a good simulation; they use sophisticated turbulence closure models that are computationally intensive and that require considerably more information than what is typically available during oil-spill simulations. Fortunately, the need for mixing parameterizations in oil-spill modeling is fundamentally different than in numerical ocean modeling and is not nearly as consequential. In oil-spill modeling, the diffusion equation is used to simulate the small-scale spreading of oil caused by oceanographic processes that are not resolved by the velocity in equation \eqref{eq:adv}.

In oil-spill modeling, the most efficient way to determine the eddy diffusion coefficient is to choose a constant coefficient that matches the observed spread of oil. This is a strategy used for hindcasts and forecasts where the main objective for the simulation of diffusion is matching the spread of oil as observed through overflights, ships, and satellites. However, during response forecasts, the diffusion coefficient is chosen to err on the high side, so that the simulations are unlikely to underestimate the extent of impacted locations.  For example, standard practice for NOAA's Office of Response and Restoration is to use a random walk with a constant diffusion coefficient as described.  Despite its rather crude and ad hoc nature, it is often important to simulate diffusive processes in oil-spill modeling because 1) it is needed to match observed oil spreading about the trajectories resulting from equation \eqref{eq:adv}, 2) it provides a least-regret conservative estimate for the spreading and impact of oil during response forecasts,  and 3) for simulations without wind, it provides one way for oil to beach which is otherwise not available from most ocean-current velocity products (e.g. ocean models set to zero the velocity normal to the coast near the coastline, although it may cause beaching if there is a mismatch between the velocity product and the model coastline, or due to numerical instabilities). Beaching due to diffusion is most appropriate in the surf zone, the effects of which are not typically simulated in ocean circulation models.  Other (more realistic) mechanisms that may drive oil beaching are discussed in section \ref{sec:four}.

There are types of oil-spill simulations that may need an automated method of determining an eddy diffusion coefficient. For example, ensemble-type modeling uses a large number of oil-spill simulations to evaluate the environmental impact of an oil spill in a statistical sense; such simulations may choose to include diffusive processes. The solution for oil-spill simulations that compute an eddy diffusivity as part of the problem (instead of choosing to match observed spread) is described in  \ref{app:diff}. Further discussion on stochastic parametrizations of diffusion and their implementation can be found in the technical documentation for NOAA's GNOME model \citep{Zelenke2012} and  \cite{Duran2016}.\\

The easiest way to numerically solve equation \eqref{eq:adv-diff} for oil transport simulations, is to separately solve \eqref{eq:adv} and \eqref{eq:diff} and then add the motion induced by each process to obtain the oil's movement. Solutions to \eqref{eq:adv} and \eqref{eq:diff} are often found separately in Lagrangian terms, i.e. by computing oil-parcel trajectories, rather than in Eulerian terms where the equations are solved over a fixed numerical grid. Because the advection part of transport preserves concentrations, the simulation of advection reduces to integrating the velocity field to obtain trajectories given an initial time and position. Thus, the typical approach to simulate the advective part of oil's trajectory is the solution $\ve{x}(t)$ to the equation:
\begin{equation}
\dd{\ve{x}}{t}= \ve{u}\left(\ve{x}\left(t\right),t\right) \quad \ve{x}(0)=\ve{x_0}.
\label{eq:ode}
\end{equation}
Equation \eqref{eq:ode} can be solved with regular numerical methods for ordinary differential equations, although there are some specific considerations due to the discrete nature of the velocity data being integrated \citep[e.g.][]{nordam2018numerical, Nordam2020}.

The diffusion part is typically modeled as a random walk, numerically simulating the random motions induced by the modeled collisions with equation \eqref{eq:diff}.   The resulting motion is added to the solution of \eqref{eq:ode} at each time step. Typically, the precision needed for $\ve{u}$ in \eqref{eq:ode} is more consequential than the spread of oil modeled with diffusion. Because of this, in this chapter, we will focus on the advective part of oil's transport. In this approach, once the advection part is satisfactory, further simulations will add diffusion to the advective part. We note that formally, the interplay between advection and diffusion is more complicated than often appreciated; the effect of diffusion in the context of Lagrangian transport is currently being researched \citep{Haller2018}.

The Eulerian representation of fluid flow and its associated numerical representation could, in principle, be used for oil-spill modeling. In practice, however, it is much more efficient to simulate oil transport using the Lagrangian representation. Among the problems inherent to the Eulerian representation is the need to set up a numerical grid for each domain.  Also, the Eulerian representation is more computationally intensive because it requires solving the advection-diffusion equation at each point on the grid, regardless of whether there is oil there or not, while in the Lagrangian approach trajectories are integrated only for existing oil parcels. In the Eulerian approach, the advection term in \eqref{eq:adv-diff} can be problematic for numerical methods  \cite[e.g.][sections 3.3, 3.4, 3.5.1, 3.5.2, 5.10]{Durran2010}, while the Lagrangian approach only requires integrating an ordinary differential equation which is, for the most part, straightforward and highly accurate. Finally, the Lagrangian approach works well with a velocity field saved at a series of discrete times, this is convenient because it readily allows additional experiments using the same velocity field. A comparison of advection results from Eulerian and Lagrangian formulations can be found in \cite{Wagner2019}.

Oil-spill models can account for other processes related to the fate of oil separately. For example, oil droplets breaking into smaller-size droplets, oil dissolution at depth, or oil evaporation at the sea surface. These processes can affect the oil's buoyancy. As oil-spill and blowout models increase in complexity, the effects of vertical motion will be included. Oil's vertical motion may be induced by ambient conditions such as waves, or by oil's buoyancy (some oils are denser than water, some are less dense), droplet size, and weathering. Further details on some of the processes causing vertical motion can be found in \cite{Nordam2020book}.

In this chapter, we assume that oil's vertical location is known, whether varying or fixed. This is a valid approach because:
1) For some spills, the oil floating at the surface is of primary concern, and thus modeling horizontal trajectories alone may be enough to get satisfactory results.
2) When vertical motion is important, vertical and horizontal components of a parcel's trajectory are computed separately. This is because the mechanisms forcing horizontal motion, tend to be different from mechanisms forcing vertical motion, and the resulting trajectory components can be added to give an updated location for the next integration step. Whenever the vertical dimension is included, the problem of determining oil's horizontal motion is more complicated. The vertical position of oil must first be determined to be able to sample the correct horizontal velocity, additionally, the velocity $\ve{u}$ driving horizontal motion will now need to be accurate at different vertical levels. 

For this chapter's discussion, horizontal transport will be defined as the motion of oil at a constant depth, whether at the surface or deeper down within the water column. Horizontal motion in this chapter also means motion along surfaces of constant density (lateral motion in oceanographic terminology) as long as it does not involve abrupt vertical displacements, e.g. where constant density surfaces rise to intersect the ocean surface. Some comments will be made regarding horizontal transport of oil at density fronts in subsection \ref{sec:fronts}. Ocean currents are driven by a variety of physical processes that, for oil-spill modeling, can be conveniently classified by their vertical location within the water column. The range of physical processes driving motion can then be narrowed down according to the vertical location of oil.

\section{Overview of oil transport in the ocean}
\label{sec:three}

In practice, transport of oil in the ocean is successfully simulated as the movement of parcels moving with the velocity $\ve{u}$ of ocean currents, as in equation \eqref{eq:ode}, this is demonstrated, for example, in \cite{Jones2016}. Near the surface, additional movement induced by wind and waves may help drive motion. In general, the vertical location of oil is consequential because the effect of wind and wave changes abruptly in the upper meters of the ocean, as will be discussed in section \ref{sec:four}, and because ocean currents also tend to change with depth.

Over the last several decades we have come to understand the ocean as a turbulent fluid in perpetual motion, rich in temporal variability. There are different types of turbulence in the ocean and a vast variety of instabilities---processes that can trigger oscillations with speeds considerably larger than the mean flow. Some of these oscillations (e.g. eddies) often result in hyperbolic points in the velocity field. Intuitively, hyperbolic points in the Eulerian velocity field suggest that initially-close trajectories are likely to undergo exponential separation. While this basic intuitive idea turns out to be correct, the relation between the velocity at each point and the trajectories traversing this time-dependent velocity is not as intuitive, requiring a careful mathematical treatment to uncover, as discussed in section \ref{sec:seven}.

\begin{figure}[H]
	\begin{center}
\includegraphics[scale=0.275]{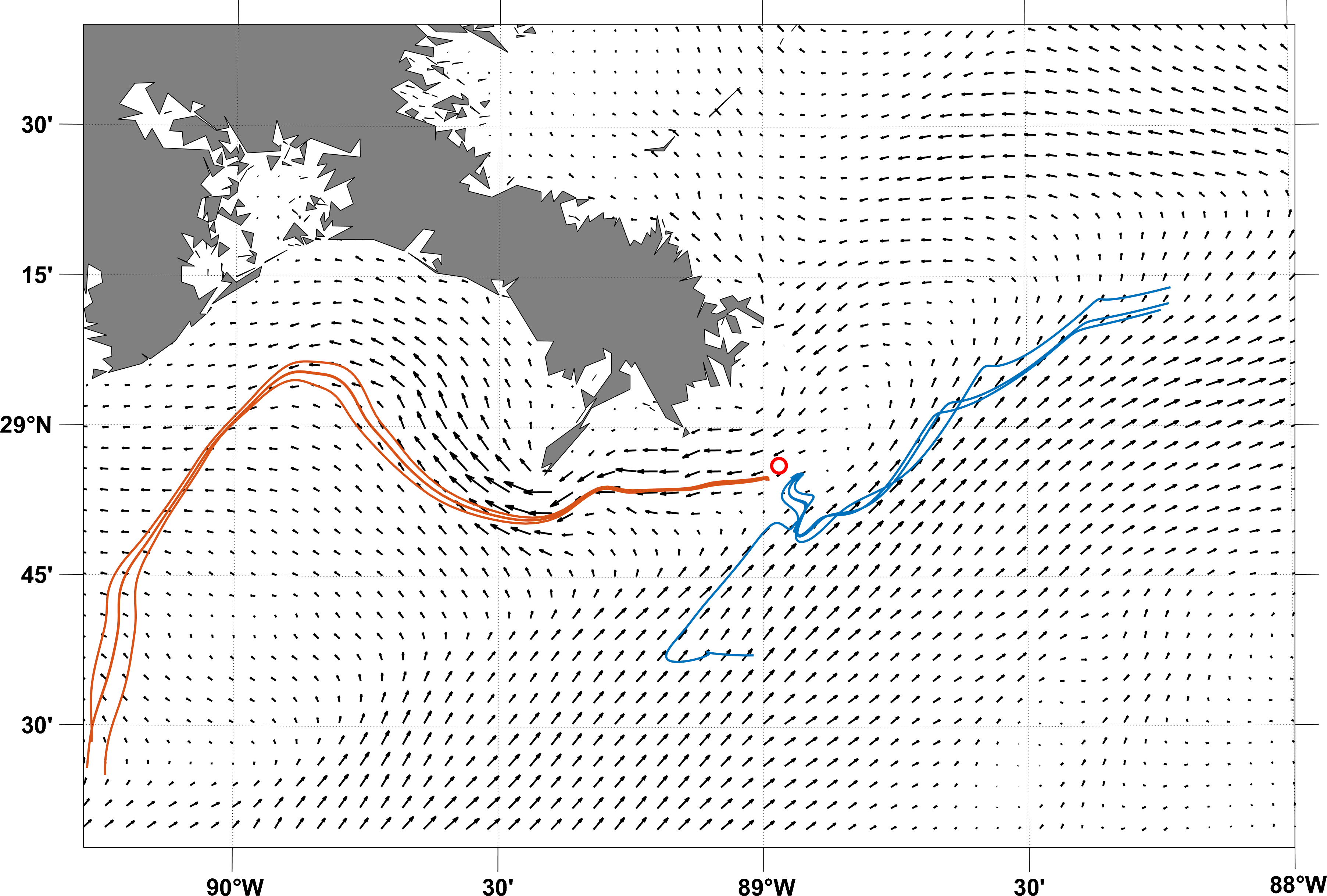}
	\caption{ Sea-surface velocity (black vectors) near the Mississippi delta in the Gulf of Mexico from a HyCOM Gulf of Mexico operational simulation on March 14, 2016. Four trajectories (orange) are released at noon March 9, 2016, just southwest of Taylor well (red circle). Another four trajectories (blue) are released just southeast of Taylor well at the same time. A hyperbolic behavior separates initially-close ($<5$km) blue and orange trajectories more than 200 km over 5 days, some moving northeast, some moving southwest. The blue trajectories were initially 300–500 m apart, yet three remain close ($<5$ km), and one separates about 100 km.  }
	\label{fig:hyp} %date of currents datestr(t(end-10))='14-Mar-2016 06:00:00'
	\end{center}  %date trajectories: '09-Mar-2016 12:00:00'    '14-Mar-2016 12:00:00'
\end{figure}

As an example to illustrate chaotic behavior, we use HyCOM-GoM (Hybrid Coordinate Ocean Model-Gulf of Mexico), a state-of-science ocean model that is likely to be used by oil-spill modelers in the Gulf of Mexico, to advect two groups of four trajectories initiated less than 5 km apart. Within each group, four trajectories are initiated within 300--500 m of each other. The two groups of trajectories undergo exponential separation as they move with realistic ocean currents, ending 200km apart after 5 days (Fig. \ref{fig:hyp}). These trajectories are initiated close to the Taylor Energy well that continued for many years to spill oil from the seafloor starting in 2004 \citep{Sun2018}, representing a realistic example of the uncertainty that an oil-spill modeler might face.  Among the group of four trajectories initiated to the southeast of Taylor well, one trajectory separates 100 km from the other three trajectories, despite being initially 300–500 m away, further showing the chaotic nature of trajectories in the ocean. This example shows that for a time-dependent flow, the interaction of trajectories with a hyperbolic point in the velocity is complex.

Exponential separation of initially close trajectories (i.e.  a sign of chaotic behavior) is an important part of why predicting trajectories in the ocean is difficult. Small errors in the velocity field or in the location at which trajectories are initiated are likely to result in large errors over short periods of hours to days. This is a problem inherent to ocean currents, it cannot be corrected with higher-order integration when computing trajectories.  We return to this type of problem in section \ref{sec:seven}.

A different type of problem occurs when a trajectory is deficient because the velocity is not representative of all the processes driving motion. In this case, it is sometimes possible to parameterize missing physics to complement the velocity. Oil does not necessarily remain at the sea surface. The depth at which oil is located is particularly important when parameterizing missing physics, as driving mechanisms change drastically even within tens of centimeters in the upper ocean, as discussed in section \ref{sec:four}.

\subsection{Sources of velocity for oil-spill modeling}

In this section, we provide an overview of some common sources of ocean current data for oil spill modeling. Two products are from remotely-sensed measurements and are therefore limited to producing a sea-surface velocity up to the current time. Ocean models produce ocean currents from numerical simulations and can provide a forecast into the future, as well as a complete velocity field, both horizontally and vertically throughout the water column.

 \subsubsection{Numerical ocean models}

The equations governing geophysical fluid dynamics---fluid dynamics on a rotating sphere---can be discretized and solved numerically. The equations themselves are very complicated, and their numerical solution is further complicated because motion at different spatial scales, from thousands of kilometers to centimeters, interact with each other in fundamental ways, yet computers are not powerful enough to simulate all such scales. Also, simulations are necessarily initiated from imperfect initial and boundary conditions, and geophysical flows tend to be chaotic. Notwithstanding, ocean models are surprisingly accurate in portraying a variety of physical processes in the ocean, and are often used as the source for the velocity $\ve{u}$ needed to solve equation \eqref{eq:ode}.

When using an ocean model, the vertical resolution should be a concern even when simulating oil transport exclusively at the sea surface. This is because the model's output for a surface velocity will be in reality a representation of the vertically sheared currents over the height of the top grid cell of the model, not a representation of the velocity at the very surface. The model output that is closest to the surface is a depth-averaged value, where averaging takes place over the thickness of the model's upper vertical level. Naturally, coarse resolutions result in greater smoothing, and therefore a less realistic representation of the surface velocity. In addition, even if a model uses a reactively thin surface layer, it usually does not include smaller-scale processes at the surface, notably wind waves.

%models may create variability that is not in the ocean
Producing accurate velocity products with models is also complicated because hyperbolic trajectories are often linked to ocean instabilities that are not completely understood and are difficult to accurately simulate. For example, ocean eddies can have a profound effect on Lagrangian transport, and a numerical simulation may develop an eddy that does not exist in the ocean. Even if a model accurately simulates an eddy, small displacements in the eddy location relative to the correct position of the eddy in the ocean can result in large trajectory errors.  Thus, ocean models often do not represent the oceanic structures that are most influential on trajectories with enough accuracy. This is true even when the numerical model assimilates a variety of ocean measurements in an attempt to replicate the real ocean. Observations used by data-assimilating models include satellite products such as sea-surface height, temperature and salinity, and in situ observations from oceanographic buoys, drifters, gliders, and other autonomous platforms. A recent overview of progress and challenges in ocean modeling can be found in \cite{Fox-Kemper2019}.

 \subsubsection{High-Frequency radars}
High-frequency (HF) radars can measure sea-surface currents remotely near the coastline ($<200$ km) with resolutions typically 500 m to 6 km, and up to hourly in time. Velocity from HF radar is  an exponentially-weighted vertical average, with a decay scale that is proportional to the wavelength of the radar signal  \citep[e.g.][]{Rohrs2015c}. Due to the resolution, which is unable to determine small-scale structures, and processing errors, trajectories from drifters designed to sample similar ocean currents as those measured by HF radar, differ from trajectories computed from HF radar velocity. Carefully calibrated radars at resolutions higher than about 1.5 km and 3 hours can replicate drifter trajectories with a separation rate of about a few kilometers over a day \citep{Rypina2014,Kirincich2012}. However, the quality of HF radar processing and HF radar resolution varies. Improving HF radar velocity products to better represent coastal currents is an ongoing endeavor \citep[e.g][]{Kirincich2019}. Other limitations include accuracy that varies with position relative the the antennae, and gaps between stations and very near the coastline.

HF radars are an important part of operational models that assimilate the surface velocity to minimize the model's error. HF radar can measure currents near the coast, making them an excellent complement to altimetry that can only produce a geostrophic velocity (section \ref{sec:alt}) further from the coast.

There are methods to improve HF radar data for Lagrangian purposes. For example, to produce trajectories from HF radar that are closer to drifter trajectories, the Eulerian velocity may be corrected using trajectory data \citep[e.g.][]{Berta2014}. A downside of this approach is that it requires deploying drifters and allowing them to drift for some time before the corrections can be applied.

HF radar is increasingly available in the U.S. and around the world \citep{Roarty2019}; a review on HF radar can be found in \cite{Paduan2013}.

 \subsubsection{Velocity products from satellites}
\label{sec:alt}
Satellites with altimeters are able to measure sea-surface height (SSH) with enough accuracy that a geostrophic velocity proportional to the SSH gradient can be computed:

\begin{align}
u_\mathrm{g}(x,y,t)&=-\frac{g}{f}\pp{\eta(x,y,t)}{y} \label{eq:alt1} \\[0.1 in]
v_\mathrm{g}(x,y,t)&=+\frac{g}{f}\pp{\eta(x,y,t)}{x} \label{eq:alt2}
\end{align}

where $(u_\mathrm{g},v_\mathrm{g})$ are respectively the east and north components of the geostrophic surface velocity, $\eta$ is the SSH, $g$ is the acceleration of gravity and $f = 2\Omega \sin \theta$ is the Coriolis parameter,  $\Omega = \SI{7.29e-5}{\per\second}$ is the rotation of the Earth, and $\theta$ is the latitude.

Velocity from altimetry has been shown to give good results for Lagrangian transport applications and may give superior results than numerical models that assimilate the same altimetry data. We cite a few studies as examples where using altimetric velocity for Lagrangian transport applications has been shown to be a good choice.

\cite{Ohlmann2001} compares surface velocity from drifters and altimetry and finds very good agreement in the Gulf of Mexico deeper than the 2000-m isobath and good agreement between the 200- and 2000-m isobaths. They find that these correlations depend on the length scale over which the differentiation in equations \eqref{eq:alt1} \eqref{eq:alt2} is computed, with best results at $\partial x , \partial y \sim 125$km. A combined observational and modeling study in northern Norway found that the accuracy of trajectories calculated from satellite products was comparable to ocean model data, and in some cases, better \citep{dagestad2019prediction}; the superior results they report from a free-running model are not surprising in coastal areas were geophysical flows are more predictable \citep[e.g.][]{Kim2011}, and altimetry measurements less reliable.

\cite{Sudre2013} show that globally the velocity from ARGO floats correlates well with the velocity they produce mainly from altimetry, except near the Equator for the meridional component. They also show excellent correlations between their altimetry-based velocity and the velocity from drifters in the Indian Ocean. \cite{Sudre2013} also show that their velocity product is capable of explaining Lagrangian transport visualized through satellite-sensed chlorophyll during an iron-release fertilization experiment.  \cite{Olascoaga2013} shows a very good correspondence between altimetric hyperbolic Lagrangian Coherent Structures (LCS) and satellite-observed transport (chlorophyll). \cite{Jacobs2014} then compares this transport to LCS from operational ocean models NCOM (Navy Coastal Ocean Model)  and HyCOM. They find that altimetry gives accurate transport patterns, while the two models show a fictitious transport barrier that is crossed by offshore chlorophyll advection (see their figures 1 and 2); they propose a modification to the data assimilation scheme as a correction. NCOM and HyCOM are the former and current models used by the U.S. Navy for their Global Ocean Forecast System that assimilates a variety of observations through the Navy Coupled Ocean Data Assimilation (NCODA) system. The results of \cite{Olascoaga2013} and \cite{Jacobs2014}  are for July 2012, here we will analyze a similar transport pattern during the Deepwater Horizon accident in May 2010 in section \ref{sec:example}. \cite{Liu2014} found that different altimetry velocity products perform similarly and that trajectories simulated from altimetry perform better than from data-assimilative models. \cite{Berta2015} compare satellite-tracked drifters to synthetic trajectories from altimetry finding ``satisfactory average results''; they also show how blending drifter data into the altimetric velocity considerably improves trajectory hindcasts, and restores missing physics that cannot be explained by Ekman superposition.  \cite{Beron-Vera2013} and \cite{Beron-Vera2018} show the relevance of Lagrangian coherence associated with eddies detected objectively from altimetry. \cite{Essink2019} found that trajectories advected with  altimetry do well in replicating the main transport patterns observed with drifters, and even though they find that a variety of statistics from altimetry trajectories do not closely resemble those from observed trajectories, we note that the concern for oil-spill modelers is identifying prevailing oil movement. Another limitation of Satellite-derived products is the low frequency of satellite passes over a given region---a synoptic view cannot be instantly obtained.

%In section \ref{sec:example}, we show that while the velocity from altimetry may still contain valuable information even when having errors.

Work towards remotely-sensed velocity products of higher resolution is currently underway \citep[e.g.][]{Chelton2019}.

% reviewed up to here
\section{Transport in the upper layer of the ocean.}
\label{sec:four}

Among the challenges oil-spill modelers face is that some of the physical processes driving oil's motion near the ocean surface may not be represented in available velocity products.  In this section, we discuss how motion in the upper layer of the ocean may be strongly influenced by wind drift (windage) and Stokes drift from waves.  As we will see, motion in the upper centimeter of the ocean can be significantly different than in the upper meter. Velocity from HF radar and ocean models do not typically include Stokes drift or windage. See \cite{Rohrs2015c} for a discussion on Stokes drift in HF radar ocean currents. Velocity from altimetry does not include Stokes drift, windage or Ekman transport, although the latter is sometimes added from additional satellite measurements.  Because it is difficult to obtain a velocity that is representative of the upper centimeter, it may be necessary to parameterize certain types of physics if the trajectories of interest are in the order one-centimeter upper layer of the ocean (or smaller: a typical oil "slick" may be on the order of microns thick). We note that near the ocean's surface, the vertical location of oil makes a big difference; even when oil has a surface expression, large amounts of oil may circulate beneath the upper centimeter. The vertical location of oil may be determined by oil's density (it can be heavier than water) or due to a dynamic balance between entrainment, vertical mixing, and rise due to buoyancy  \citep{Nordam2020book}.  %For further details, see the vertical chapter, and, e.g., \cite{elliott1986, johansen1982, nordam2019naive}.

Only recently have adequate observations resulted in insight into the movement within the upper layer of the ocean. It is therefore timely to review how this information is relevant for oil spill modelers, as it suggests possibilities that were not typically considered in the past.  We include the Point Wells spill in the Salish Sea as a recent example where this type of physics was used to explain the oil's trajectory after remaining a mystery since 2003. 
Cross-shelf transport is crucial for oil-spill modeling because it is needed for oil to beach, and beaching is one of the more consequential events during oil spills.  A large fraction of ocean currents is in approximate geostrophic balance, strongly constraining flow in its ability to cross isobaths \citep[][]{Brink2016}. Consequently, Lagrangian transport near the coast tends to move parallel to the coastline (more precisely along geostrophic contours) and is limited in its ability to move perpendicular to the coast  \citep[][]{LaCasce2008}. Among the processes capable of causing cross-shelf transport are eddying activity, ageostrophic processes such as Ekman transport, Stokes drift, and windage. We also present recent evidence that windage and Stokes drift are important because they are effective in driving large-scale beaching.
%
%Eddies can also be very efficient drivers of cross-shelf motion, albeit more localized. A dramatic example (velocity reaches almost 1.8 m/s, and cross-shelf transport reaches 16 Sv) can be found in \cite{Malan2020}. This means that the oil-spill modeler needs to watch out for eddies near the coast, do they really exist or is it only ocean-model variability? If they do exist in reality and the model is reproducing an eddy causing cross-shelf transport, is the location of the eddy accurate? The answer to these type of questions should help identify whether there is spurious transport and whether the model solution is adequate for that particular hindcast/forecast.\\

\subsection{Windage}
\label{windage}

For some time now, it has been noted that as the wind increased in magnitude, the effect it had on motion near the sea surface increased. Recently, \cite{Lodise2019} used data from one of the largest Lagrangian experiments to date, to show that undrogued drifters sampling the upper 5 cm of the ocean move with a velocity that is 3.4–6.0 \% of the wind under strong wind  (12–20 m/s) conditions, with a deflection to the right of wind direction increasing from 5 to 55\dg, as wind increased from 12 to 20 m/s. For drogued drifters sampling the upper 60 cm of the ocean, the angle of deflection increased with increasing wind (12--20 m/s) from  30 to 85\dg, and windage ranged between 2.3–4.1\% of wind speed. In those experiments, an additional velocity component from Stokes drift was found to be about 1.2–1.6\% of the wind for undrogued drifters and about 0.5–1.2\% of the wind for drogued drifters, with a deflection to the left from wind's direction of about 5\dg.  Overall, windage and Stokes drift accounted for about 70\% of the total velocity of drogued drifters, and about 80\% of the velocity of undrogued drifters.   \cite{Laxague2018} use a variety of instruments including different drifters to measure currents in the upper meters of the ocean with an emphasis on the upper centimeters. They find that the velocity in the upper centimeter, about 60 cm/s, is twice the velocity averaged over the upper meter, and four times the velocity averaged over the upper 10 meters. \cite{Rohrs2015b} use two types of drifters, an undrogued drifter to sample the surface layer and a drogued drifter to sample the upper 70 cm. They find that the upper layer is influenced by wind, while subsurface motion has a stronger link to ocean dynamics, the result being that the surface response to wind forcing is distinct from the response 70 cm below. \cite{Androulidakis2018} also show examples of how drogued drifters have significantly different trajectories than undrogued drifters, with the latter heavily influenced by wind.

Often the direction of motion induced by wind is not the same as the wind direction. However, the angle of deflection can vary with several factors including the ocean's stratification, the buoyancy of the particle, latitude, and the magnitude of the wind, making it difficult to predict the direction of windage. Another complication with oil spills is that the windage of the oil changes over time as the oil weathers and is transformed. Similar to the strategy where the eddy diffusion coefficient is adjusted to match the observed spread of oil, it might be necessary to use observations when possible, to adjust the windage coefficient and the angle of deflection to match the observed motion. Some discussion on the diverse range of deflection angles that have been measured at sea can be found in \cite{Duran2016}; windage for different objects can be found in \cite{breivik2011wind,maximenko2018numerical}. %I wanted to keep at observations dedominicis2013medslik, }.

\subsubsection{The Point Wells oil spill.}

On midnight of December 30, 2003,  almost 5,000 gallons (about 110 barrels) of fuel spilled into the Puget Sound after a tank barge accidentally overtopped near Richmond Beach in Shoreline, Washington. Helicopter overflights early in the morning observed that the surface expression of the spill drifted south, yet oil-spill models forecasted northward movement. The temporal and spatial extent of the spill was small enough (about a day and a 15 km trajectory) for overflights to suffice supporting response efforts. However, the reason why typical oil-spill model forcing, such as ocean currents and 1--3\% of the wind, could not explain the surface-oil trajectory remained a mystery. This was particularly puzzling in an enclosed sea where predictable tides are an important component of the circulation. \\

Published work where windage reaches 6\% of the wind is unusual in the oil spill modeling community, although not unprecedented. In \cite{Duran2018BLOSOM} it was conjectured that oil's motion during the Point Wells oil spill was driven by a combination of 6\% of the wind with a 9$^\circ$ deflection to the right of wind's direction and ocean currents.  This hypothesis explained the trajectory of the spill that had previously been a mystery, although the use of such a high windage coefficient was unusual. It wasn't until a year later that measurements of motion in a fine upper layer of the ocean were published by \cite{Lodise2019}, documenting motion dominated by windage at 6\% of the wind speed.

\begin{figure}[H]
	\begin{center}
\includegraphics[scale=0.45]{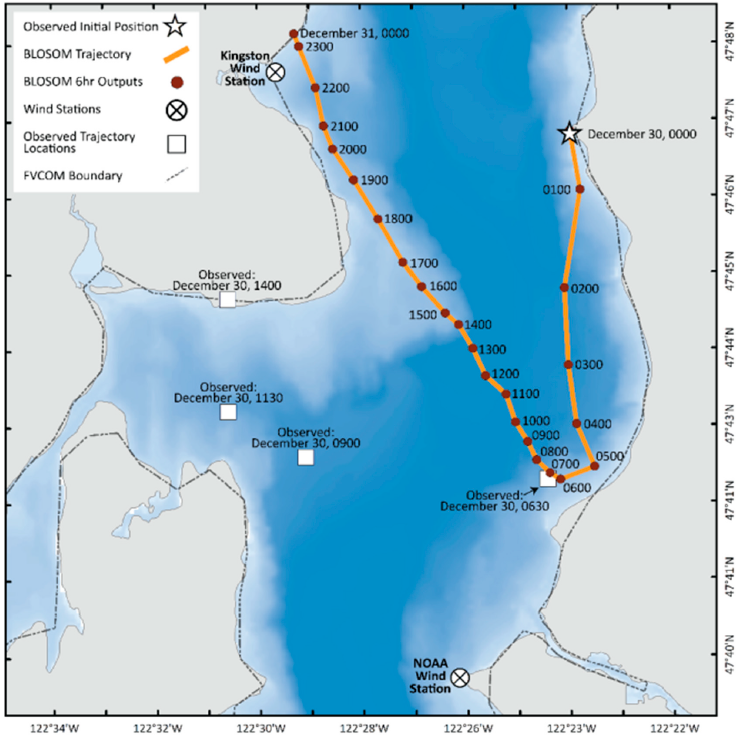}
	\caption{Trajectory (orange, red circles mark locations at hourly intervals) initiated at the time and location of the Point Wells 2003 spill in the Salish sea (white star) forced only with 6\% of the wind measured by the NOAA wind station (white circle with black cross near bottom). Locations and times where oil was observed  are marked with white squares and text.  }
	\label{fig:wind}
	\end{center}
\end{figure}

Hindcasts of the Point Wells spill advecting oil only with wind from a meteorological  station in the vicinity of the spill were suggestive for two reasons: 1) excellent agreement with the observed trajectory in the first six hours and 2) it forced oil towards the south, an observed-trajectory feature that had been difficult to emulate with ocean currents and typical windage (Fig. \ref{fig:wind}). When 6\% of the wind from a meteorological station, and ocean currents from an ocean model that replicated tides with high skill, were combined to force the trajectory computation, the resulting trajectory matched the correct locations at the correct times throughout the spill, finally beaching at the correct location in the afternoon of December 30, 2003. A full account of the numerical experiments can be found in \cite{Duran2018BLOSOM}.

\subsection{Stokes drift}
Stokes drift is a net drift in the direction of wave propagation caused by the asymmetrical orbital motion of particles near the surface induced by passing waves. Some authors convey the idea that Stokes drift is canceled in the mean due to the Coriolis effect. However, there is a large body of evidence suggesting that cancellation in the near-surface is negligible in the presence of turbulence induced by wind stress, which is the typical condition in the ocean   \citep[e.g.][]{Clarke2018}. Stoke's drift was an important driver of oil during the Deepwater Horizon: it was responsible for the observed beaching patterns and it is believed to have avoided oil being entrained by the Loop Current \citep{Carratelli2011,LeHenaff2012,Weisberg2017}.  This is consistent with other studies reporting that  Stokes drift can exceed the Eulerian mean in the cross-shelf direction  \citep[e.g.][]{Monismith2004}. 

Stokes drift is mainly driven by high-frequency waves, that is, waves forced by local wind rather than remote swell \citep{DAsaro2014,Clarke2018}. Using many years of hourly concurrent wind and directional wave spectra from buoys in the Gulf of Mexico and the Pacific, \cite{Clarke2018} derived a simple formula with which Stokes drift can be parametrized directly from local wind, with good accuracy (within about 1 cm/s from the average Stokes drift) for common wind speeds (between 1 to 50 m/s):

\begin{equation}
u_{\mathrm{Stokes}}=4.4 u_* \ln\left(0.0074 U_{10}/u_* \right)
\end{equation}
where $u_{\mathrm{Stokes}}$ is the magnitude of Stokes drift, $U_{10}$ is the wind speed 10 meters above sea level and $u_*=\sqrt{\tau/\rho_0}$  is the frictional velocity, the square root of wind-stress magnitude divided by a reference seawater density.  The direction of Stokes drift is given by the unit vector in the direction of the wind. This is good news because local wind data is often available, whether from meteorological stations or operational models. Further good news is that this result holds even in the presence of swell.

%The magnitude of the wind stress, $\tau$, was parameterised as \cite{Clarke2018}
%%
%\begin{subequations}
%\begin{align}
%    \label{eq:tau}
%    \tau &= \rho_{air} C_D U^2_{10}, \\
%    \label{eq:CD}
%    1000 C_D &= 2.70/U_{10} + 0.142 + 0.0764 \cdot U_{10},
%\end{align}
%\end{subequations}
%%
%where $\rho_{air}$ is the density of the air, $U_{10}$ is given in units m/s, and the numerical prefactors presumably have the required units to render $C_D$ dimensionless.

Both \cite{Clarke2018} and \cite{Onink2019}  note that an additional contribution to transport at the surface, also in the direction of the wind, may be necessary due to wave breaking. It is also possible that swell may induce surface transport near the coastline, as waves become increasingly nonlinear due to interactions with the bottom. Deeper down within the water column, Stokes drift from internal waves at the pycnocline has also been observed to be an effective driver of oil transport \citep{Shanks1987}. There is also a challenge when applying Stokes drift to oil transport: Stokes drift is the integrated motion over the depth of the waves. But oil generally floats, so is either at the very surface as a film, or at higher concentrations near the surface.

\subsection{Horizontal organization induced by vertical motion}
\label{sec:fronts}

How tracers respond in the upper ocean when they sample velocity structures with influential vertical motion along their path is an active topic of research. It is receiving considerable attention as new observational tools and experiments allow measuring smaller-scale processes, while higher numerical-model resolutions become accessible. Considerable progress was made when such technological advances coincided with funding that became available following the Deepwater Horizon accident. 

Identification of submesoscale structures, such as fronts and Langmuir cells, can be important because actionable countermeasures while responding to an oil spill require oil to reach a certain thickness. Intense convergence of oil along water-mass boundaries may, therefore, create an ideal location for mitigation strategies when conditions are right \citep{Gula2014}. It has often been observed during oil spills that oil collects in windrows formed by Langmuir circulation as well as at convergences associated with fresh water at river mouths. Water-mass subduction forecasting and detection is therefore suggested as an aid to identifying regions of thick oil. It should also be cautioned that using divergence as a diagnostic to identify regions of accumulation can lead to false positives and negatives \citep{Serra2020}. Clustering may happen in a region of positive Eulerian-velocity divergence, we present an example in section \ref{sec:OECS}.  Another potential caveat is the effect of strong wind acting directly on buoyant oil.  The experiment in \cite{Romero2019} suggests that wind order 10 m/s does not impede strong vertical motion at fronts, although the tracer in their experiment was neutrally buoyant, so may not reflect the behavior of a positively-buoyant tracer such as oil.

Types of vertical motion that are known to affect the horizontal distribution of oil are related to ocean    fronts, filaments, and Langmuir circulation. When two water masses meet, a front is formed along the boundary between the two. Whether one water type sinks under the other because it is heavier, or because of cabbeling, fronts in the ocean tend to be accompanied by water subduction. A vertical circulation due to similar reasons also forms along the boundaries of filaments \citep{mcwilliams2017}.  Thus, a downward vertical velocity is induced at the boundaries of water masses, which implies convergence in the horizontal plane. Frontal regions are characterized by relative vorticity and negative divergence that can be several times greater than planetary vorticity. This can have a profound local effect on transport, collapsing floating material to essentially a point. An example of drifters originally spanning a width of about 10 km, collapsing to 60 meters, can be found in \cite{Dasaro2018}. A comparison between two- and three-dimensional circulation at a scale of about 100 m, illustrating the distribution of a neutrally buoyant tracer due to vertical motion, can be found in \cite{Romero2019}, showing the tracer sinking relatively rapidly. If the tracer were buoyant, as often is oil, an agglomeration of tracer could be expected at the surface.  \cite{Androulidakis2018} found that the front-induced circulation dominated the trajectories of undrogued drifters, even under considerable wind, although wind may modulate their speed along the front. The wind is one of the factors determining the location of the front. Fronts are also of interest because they may serve as transport barriers \citep{Androulidakis2018}. 

Langmuir circulation in its most basic form results from the interaction of Stokes drift induced by surface waves, and the vertical shear induced by the turbulent transfer of momentum from wind to the upper ocean \citep{Thorpe2004,Sullivan2012}. The book by \cite{buhler2014} describes how a mean flow is induced by an instability  (Craik-Leibovich instability), a mechanism that turns out to be robust and therefore explains why Langmuir cells are ubiquitous in the ocean. As with the circulation associated with fronts, Langmuir circulation also has an important vertical component, that likewise concentrates oil into bands within minutes to hours, typically at scales of meters, to hundreds of meters \citep{Chang2019,Simecek-Beatty2017,Dasaro2000}. Convergence due to frontal circulation may dominate convergence due to Langmuir circulation \citep{Romero2019}. Changes in the vertical location of oil droplets  \cite[e.g.][]{McWilliams2000} induced by Langmuir circulation enhance the dispersion of oil by subjecting droplets to different ocean currents, as determined by vertical shear \citep{Thorpe2004}.  Also, at least sometimes, Langmuir circulation may be a more important part of ocean dynamics than previously thought \citep{DAsaro2014}.  For example, the newest ocean climate models---designed to study climate change---now parametrize the effect of Langmuir mixing at the ocean's surface. Evidence that Langmuir circulation may induce a large-scale coastal circulation can be found in \cite{Kukulka2012}.  

 In the larger picture, the spatial scales of intense subduction structures (hundreds of meters to a few kilometers) imply that they are likely to be embedded within larger ($>$50–100 km) mesoscale structures that advect the smaller structures and therefore determine their location \citep{McWilliams2019,Dasaro2018,Androulidakis2018,Jacobs2014}. This suggests that the horizontal motion might still be dominated by larger-scale features, although the local organization might be strongly influenced by the smaller scales. However, separating the length scales of oceanographic processes driving motion is not a trivial endeavor \citep[e.g.][]{Essink2019,Beron-Vera2016}. 
 
Because regions of subduction (fronts, filaments, and Langmuir cells) are ubiquitous in the ocean, it is suggested that effective oil-spill planning and response should study how to incorporate the associated material clustering in near real-time. Satellite and other remotely sensed data, such as sea-surface temperature sensed by an airplane or an UAV (Unmanned Aerial Vehicle, a drone), may be  accurate and relatively inexpensive means of identifying regions of subduction during response operations, complementing the information available from HF radar and numerical ocean models. Sea-surface velocity measured from shipboard X-Band radar seems a promising way to identify the strength of the velocity divergence \citep[e.g.][]{Lund2018}, although the reach of radar measurements shown in studies so far (less than 10 km) may be small for an appropriate sampling of the more relevant Lagrangian quantity of along-path divergence. Further research will be needed to understand the interaction between confluence unrelated to divergence, wind, and buoyant material accumulation related to vertical motion.

\section{Modern Lagrangian tools}
\label{sec:seven}
The regions where the separation of initially-close trajectories and the attraction of initially-separate trajectories occur are regions of special interest when studying horizontal motion. It is these regions that have an exceptional influence on the movement of nearby parcels, and thereby play a leading role in organizing the flow into identifiable and predictable patterns. In a two-dimensional flow, these regions are hyperbolic lines, in a three-dimensional flow they are surfaces. A rigorous approach to detecting these regions has been developed by identifying regions with maximal normal attraction, typically referred to as hyperbolic Lagrangian Coherent Structures \citep[LCS;][]{farazmand2012computing, Haller2015}.

LCS theory builds on the concept of a flow map, a function that maps every initial $(t=t_0)$ position within a domain of interest $\ve{x_0} \in U$, to its current $(t=t_1)$ position $\ve{x}(t_1)$, that is: $\ve{F}_{t_0}^{t_1}(\ve{x_0})\coloneqq \ve{x}(t_1;\ve{x_0},t_0)$.  The Jacobian of the flow map $\mathrm{D} \ve{F}_{t_0}^{t_1} $ is given by
\begin{equation}
\mathrm{D} \ve{F}^{t_1}_{t_0}(\mathbfit{x}_0) \coloneqq
 \begin{pmatrix}
   \dfrac{\partial x}{\partial x_0} & \dfrac{\partial x}{\partial y_0}\\[0.14 in]
  \dfrac{\partial y}{\partial x_0} & \dfrac{\partial y}{\partial y_0}\\
 \end{pmatrix}
 \label{eq:DF}
\end{equation}
Informally, the Jacobian of the flow map \eqref{eq:DF} can be used to map trajectory perturbations from one time to another, and this linear approximation can then be used to optimize quantities of interest. For example, normal attraction of nearby fluid parcels along a trajectory over a time interval can be maximized with respect to perturbations of the initial-time normal vector. This is the strategy used to find hyperbolic LCS, trajectories characterized by maximal normal attraction or repulsion. Working out the math for this optimization problem---a formal account of which can be found in \cite{Haller2015} and references therein---the Cauchy-Green (CG) strain tensor arises naturally. The CG tensor is defined as
\begin{equation}
  \ve{C}^{t_1}_{t_0}(\ve{x}_0) \coloneqq
   \left[\mathrm{D} \ve{F}^{t_1}_{t_0}(\ve{x}_0)\right]^\top
  \mathrm{D} \ve{F}^{t_1}_{t_0}(\ve{x}_0).
  \label{eq:C}
\end{equation}
In particular, the eigenvalues $0 < \lambda_1(\mathbfit{x}_0) < \lambda_2(\mathbfit{x}_0)$ and normalized eigenvectors $\hat{\vectorsym{\xi}}_1(\mathbfit{x}_0) \perp \vectorsym{\hat{\xi}}_2(\mathbfit{x}_0)$ of \eqref{eq:C}, are used to set up ordinary differential equations from which hyperbolic, elliptic and parabolic LCS can be found \citep{Haller2015}. Thus, the CG tensor is central in LCS theory. Note that to obtain the CG tensor one must integrate the velocity; we return to the CG tensor in section \ref{sec:OECS}. %not sure the autonomous result transfers over to arbitrary time dependance. %Different approaches to obtaining the CG tensor have been described in the literature, including numerically obtaining and differentiating the flow map \cite{farazmand2012computing}, or directly calculating the components of the tensor by a variational approach \cite[Appendix C]{oettinger2016autonomous}. We return to the CG tensor in section \ref{sec:OECS}.\\

The mathematical formality behind LCS has proven a versatile approach to understanding Lagrangian motion. Hyperbolic LCS will accurately identify how fluid will deform (i.e. along attracting hyperbolic LCS), anticipating the more influential transport patterns. However, the final results ultimately depend on the accuracy of the velocity field, which is what induces Lagrangian transport in the first place. If a velocity field is relatively accurate while having localized errors, trajectories traversing the time and location of errors in the velocity are likely to give wrong results relative to observed trajectories. LCS will be negatively affected by those velocity errors as well, correctly identifying transport induced by the velocity, yet remaining incorrect relative to observed transport. The computation of trajectories propagates localized velocity errors, often resulting in trajectories that are incorrect relative to observations when hindcasting or forecasting the transport of oil. The need to integrate the velocity field limits the suitability of observational velocity data sets and of ocean models that assimilate such data, for Lagrangian transport purposes.

Velocity products in our time can be relatively accurate thanks to relatively accurate measurements over wide areas, with satellite altimetry being particularly relevant because of good global coverage, and because it captures what is often an important part of the velocity at the sea surface (section \ref{sec:alt}). However, the coarse temporal and spatial resolution of altimetry means that the resulting velocity will almost certainly have deficiencies. Ocean models assimilating data inherit these deficiencies and have shortcomings of their own.  Given the chaotic sensitivity of trajectories, it has therefore been a natural development to try to bypass the sensitivity resulting from localized velocity errors.

Another complication with hyperbolic LCS from an applied point of view is that there is a timescale $T$ involved in the computation. When computing \eqref{eq:DF}, a choice must be made for the initial time $t_0$ and the final time $t_1=t_0+T$. The choice for $T$, sometimes even the choice for $t_0$, are often not clear a priori, forcing subjective choices. Since LCS are material lines moving with turbulent flow, these choices can result in big differences. In subsection \ref{sec:OECS} we describe a way to bypass sensitivity to the velocity field, and the need to choose $T$.

\subsection{Eulerian Coherent Structures}
\label{sec:OECS}

The fundamental equation translating from the Eulerian and Lagrangian characterizations of fluid flow is \eqref{eq:ode}, an equivalence between the velocity of a parcel traversing a trajectory $\ve{x}(t)$, and the Eulerian velocity $\ve{u}$, at the parcel's time and location. Based on this instantaneous correspondence between Lagrangian and Eulerian descriptions of fluid flow, it is natural to search for an Eulerian counterpart to the CG tensor, seeking to describe fluid deformation near the instantaneous limit. As mentioned in section \ref{sec:seven}, the CG tensor is central to finding hyperbolic LCS, trajectories that maximize normal attraction, thereby maximizing the influence on nearby water parcels and thus organizing flow. \cite{Serra2016} developed OECS, including attracting hyperbolic OECS, they showed that the Taylor expansion of the CG tensor with respect to time is given in terms of the strain-rate tensor $\ve{S}$:

\begin{equation*}
  \ve{C}^{t_1}_{t_0}(\ve{x}_0) = \ve{I} + 2 \ve{S}(\ve{x}_0,t_0)\left(t_1-t_0\right) + \mathcal{O}\left(|t_1-t_0|^2 \right). %  \label{eq:Ctaylor}
\end{equation*}

This means that for time close enough to $t_0$, Lagrangian deformation is approximated by the Eulerian strain-rate tensor, the $ij$-th entry of which is given by $\frac{1}{2}\left( \ppf{u_i}{x_j}+\ppf{u_j}{x_i} \right)$. The strain-rate tensor has eigenvalues $s_1$, $s_2$ with corresponding eigenvectors $\ve{e}_1$ and $\ve{e}_2$. Attracting hyperbolic OECS are tangent to $\ve{e}_2$, their cores given by minima in the eigenvalue $s_1$, which is the rate of change of the length of the normal eigenvector $\ve{e}_1$ due to the deformation induced by the flow; equivalently, $s_1$ is the strength of attraction normal to $\ve{e}_2$. Negative values of $s_1$ mean that there is attraction normal to $\ve{e}_2$, the more negative the stronger the attraction. Thus, \cite{Serra2016} extended the theory of LCS from finite time to their instantaneous limit in terms of an Eulerian quantity, where there is no longer a need to integrate the velocity field, effectively bypassing the attendant sensitivity.

The strain-rate tensor is objective, i.e. the results from computing Eulerian Coherent Structures are frame invariant \citep{Haller2015,Serra2016}. This means that changes of reference frames characterized by time-dependent rotations and translations will not affect the results. This is important because non-objective methods might give different results under coordinate transformations, e.g. Eulerian-velocity hyperbolic points are not Galilean invariant \citep{Serra2016}.

\begin{figure}[H]
	\begin{center}
\includegraphics[scale=0.34]{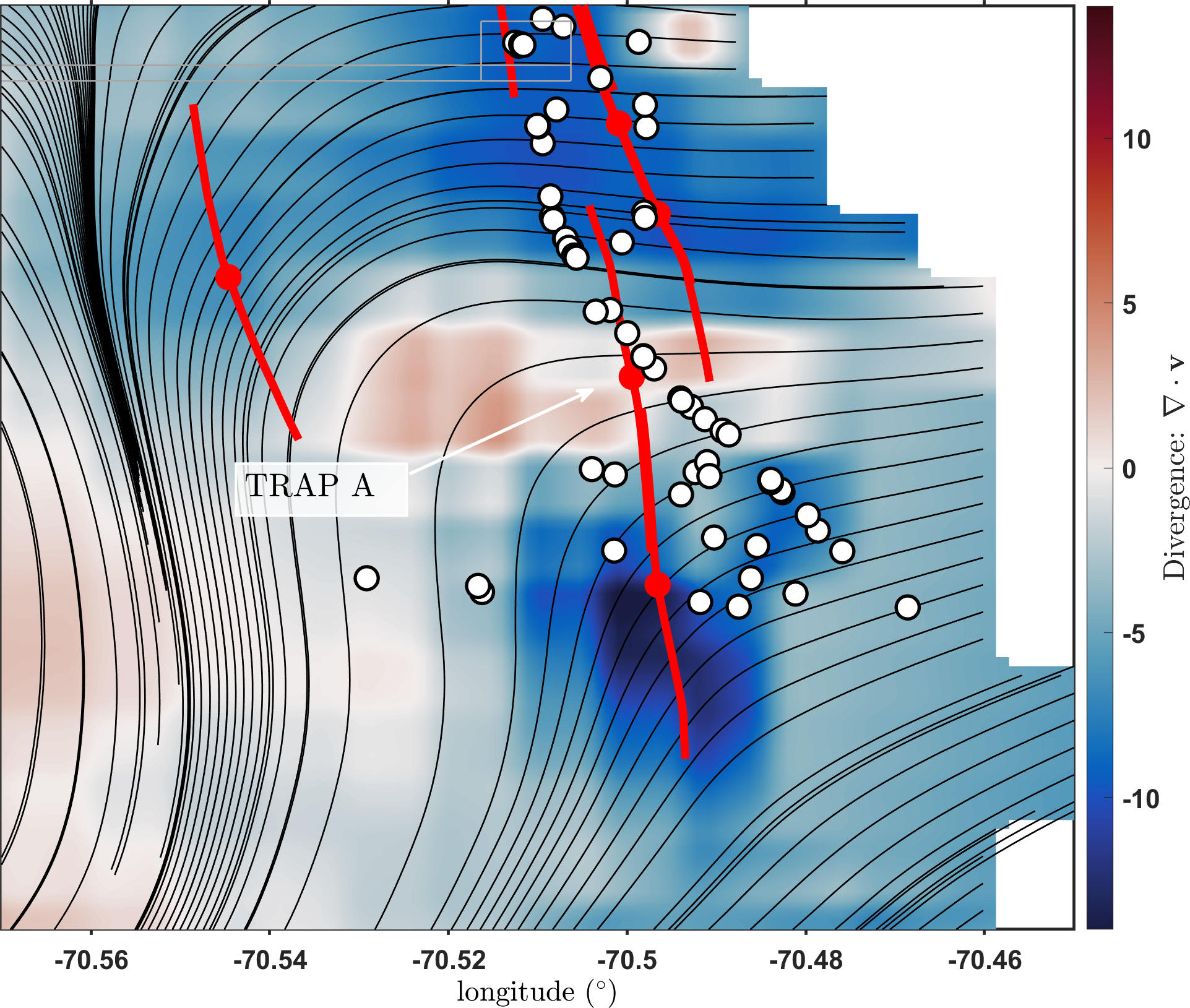}
	\caption{TRAPs (red lines) and their cores (red circles) computed from HF radar velocity off Martha's Vineyard, MA, plotted over the velocity divergence (color contours; day$^{-1}$) with satellite-tracked drifters (white dots) converging to TRAPs. Black lines are streamlines. TRAP A is in a region of positive horizontal velocity divergence. TRAPs remain invisible to divergence fields and streamlines.}
	\label{fig:traps}
	\end{center}
\end{figure}

\cite{Serra2020} use attracting hyperbolic OECS, which they call TRAPs (TRansient Attracting Profiles), to demonstrate in a series of experiments with satellite-tracked drifters and Search \& Rescue Training Manikins, that TRAPs organize flow and perform better than trajectory computations in predicting drifter locations. They use a carefully calibrated HF radar velocity and a data-assimilating model similar to what the U.S. Coast Guard would use for search and rescue operations. Similar to hyperbolic LCS, TRAPs are lines in a two-dimensional flow; they have a core which is where normal attraction is maximal, with attraction strength decaying along the rest of the TRAP. We present an example of satellite-tracked drifters converging to TRAPs computed from HF radar velocity in Martha's Vineyard in Massachusetts (Fig. \ref{fig:traps}). In this example, there is a confluence of drifters at a TRAP where the Eulerian horizontal velocity divergence is positive. A description of these data, how TRAPs organize Lagrangian motion, and how TRAPs can be used for search and rescue operations, can be found in \cite{Serra2020}. In the next section (section \ref{sec:example}), we present an example where TRAPs can predict the movement of oil at least 8 days in advance, while trajectories diverge from the observed transport due to a likely-erroneous hyperbolic point in the velocity.

\subsection{Revisiting the Deepwater Horizon with modern tools}
\label{sec:example}
The difficulty of simulating Lagrangian transport can be easily experienced by trying to replicate observed trajectories. During the Deepwater Horizon, at least six different ocean models were used in an attempt to forecast the location of oil, to provide critical information for response and planning \citep{Liu2011a}. However, there was enough intermodel variability that ensemble averaging was recommended to produce a forecast that was more likely to occur. Even then, forecasts were limited to two days due to forecast error growth. We note that these were ideal conditions as often there aren't that many ocean models available for ensemble averaging.

Here we present a different method that seeks to bypass the sensitivity that causes error growth, using TRAPs which are computed from instantaneous snapshots of an Eulerian velocity. We show that an analysis combining TRAPs and LCS is enough to 1) accurately forecast the observed movement of oil at least 8 days in advance, and 2) understand why the simulated Lagrangian transport does not conform to observations. In this example, forecasts only depend on a previous-day single velocity snapshot, and the LCS are computed from only past information to complement the information obtained from TRAPs and pinpoint the source of error in the velocity field, and its Lagrangian manifestation. %Accurate results are obtained from altimetry velocity, and from the U.S Navy operational model HyCOM Global, while HyCOM GoM at a higher resolution results in the

A variety of velocity products from altimetry are available, some of them including an Ekman component. The product that we use here is a daily velocity by GEKCO2 \citep[Geostrophic and EKman Current Observatory;][]{Sudre2013}, from satellite altimetry and wind. We confirm our results using a daily instantaneous velocity from HyCOM Global at about 9 km resolution in the Gulf of Mexico, the current U.S. Navy Operational model \citep{Burnett2014} that assimilates a variety of observations using the Navy Coupled Ocean Data Assimilation (NCODA) and that is forced with the NAVy Global Environmental Model (NAVGEM). Transport simulated with HyCOM Gulf of Mexico, which has a similar setup as HyCOM Global, but at a 4 km resolution, also produces similar simulated transport as the other velocity products used in this experiment. %, however TRAPs are less informative with the higher resolution.

We analyze daily forecasts of the movement of oil during the Deepwater Horizon accident by comparing TRAPS to the observed outline of oil and the advection of oil obtained by integrating the velocity field, i.e. computing trajectories between May 11–17, 2010, that are initiated at the location of oil observed on May 10, 2010. TRAPs are computed from snapshots of the velocity on previous days, thus providing forecasts within this hindcast exercise.

\begin{figure}[H]
	\begin{center}
\includegraphics[scale=0.37]{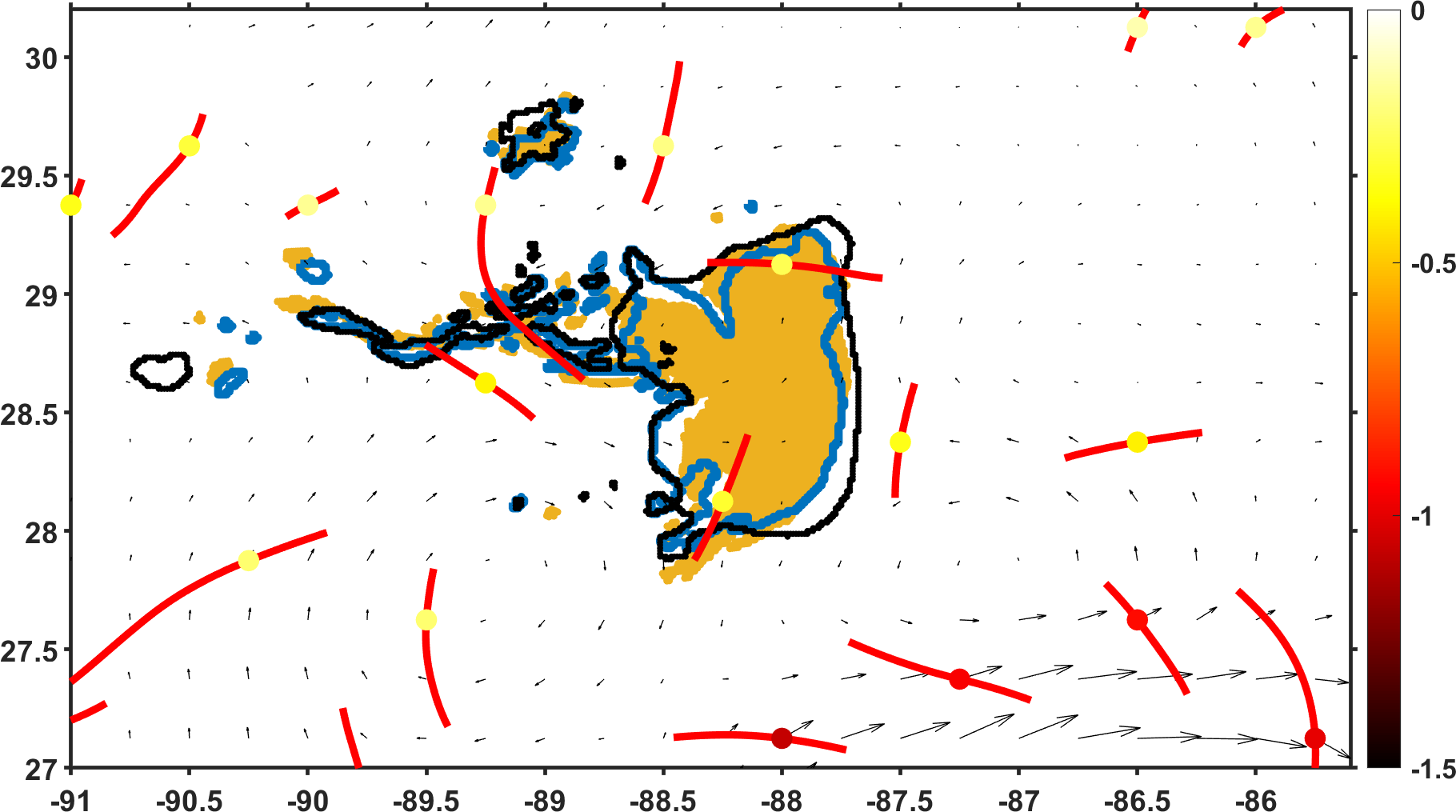}
	\caption{The blue line is the outline of oil as observed from satellites on May 10, 2010,  used as initial conditions for trajectory computations. In orange are the final positions (May 11) of trajectories initiated within the blue outline on May 10, computed by integrating GEKCO2 velocity. The black line is the outline of oil as observed on May 11, 2010.  Black vectors are the velocity from GEKCO2 on May 10, and the red lines are TRAPs computed with the velocity on May 10, TRAP cores (colored circles) are colored according to attraction strength (day$^{-1}$; colorscale on the right).}
	\label{fig:trap11}
	\end{center}
\end{figure}

The forecast on May 10 shows a weak TRAP (near 28.2N, 88.3W) suggesting slight oil movement towards the southwest, coinciding with the outline of oil observed on May 11, and with trajectories computed between May 10--11, although the simulated oil and the TRAP are slightly offset from the observed oil (Fig. \ref{fig:trap11}). The strength of attraction of the TRAP is low (about 0.3 day$^{-1}$), accurately forecasting slight oil movement.  The only other TRAP core in contact with the observed oil (near 29.2N, 88W), is the core of a TRAP almost entirely contained within the observed oil on May 11, and therefore cannot be expected to cause a significant rearrangement of oil outside of the observed oil outline. TRAPs in the southeast section of figure \ref{fig:trap11} have higher strengths of attraction of about 1 day$^{-1}$, but are still relatively far from the oil. %Other TRAPs that intersect the oil

\begin{figure}[H]
	\begin{center}
\includegraphics[scale=0.37]{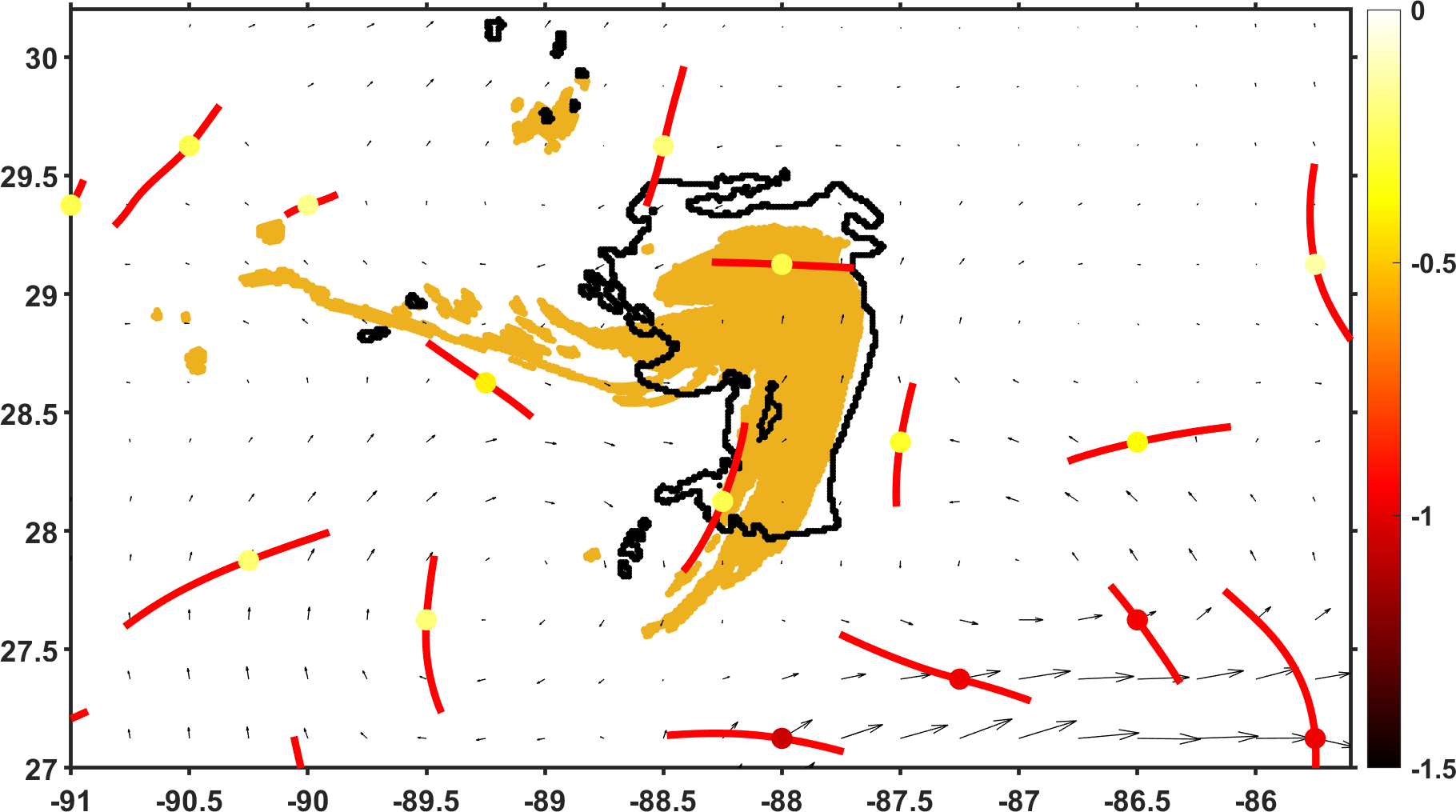}
	\caption{Same as in figure \ref{fig:trap11} but on May 13, 2010. }
	\label{fig:trap13}
	\end{center}
\end{figure}

By May 13, oil trajectories continued along the path towards the southwest, the weak TRAP computed from the May 12 velocity accurately forecasting that path (Fig. \ref{fig:trap13}). This will be the last day this weakly-attracting TRAP appears near 28.2N, 88.3W. The TRAPs with the strongest attraction computed with the May 10 velocity remain when TRAPs are computed with the May 12 velocity (Figs. \ref{fig:trap11} and \ref{fig:trap13}), these are the TRAPs in the southeast of our domain, with attraction three to four times stronger than TRAPs directly interacting with oil.

\begin{figure}[H]
	\begin{center}
\includegraphics[scale=0.37]{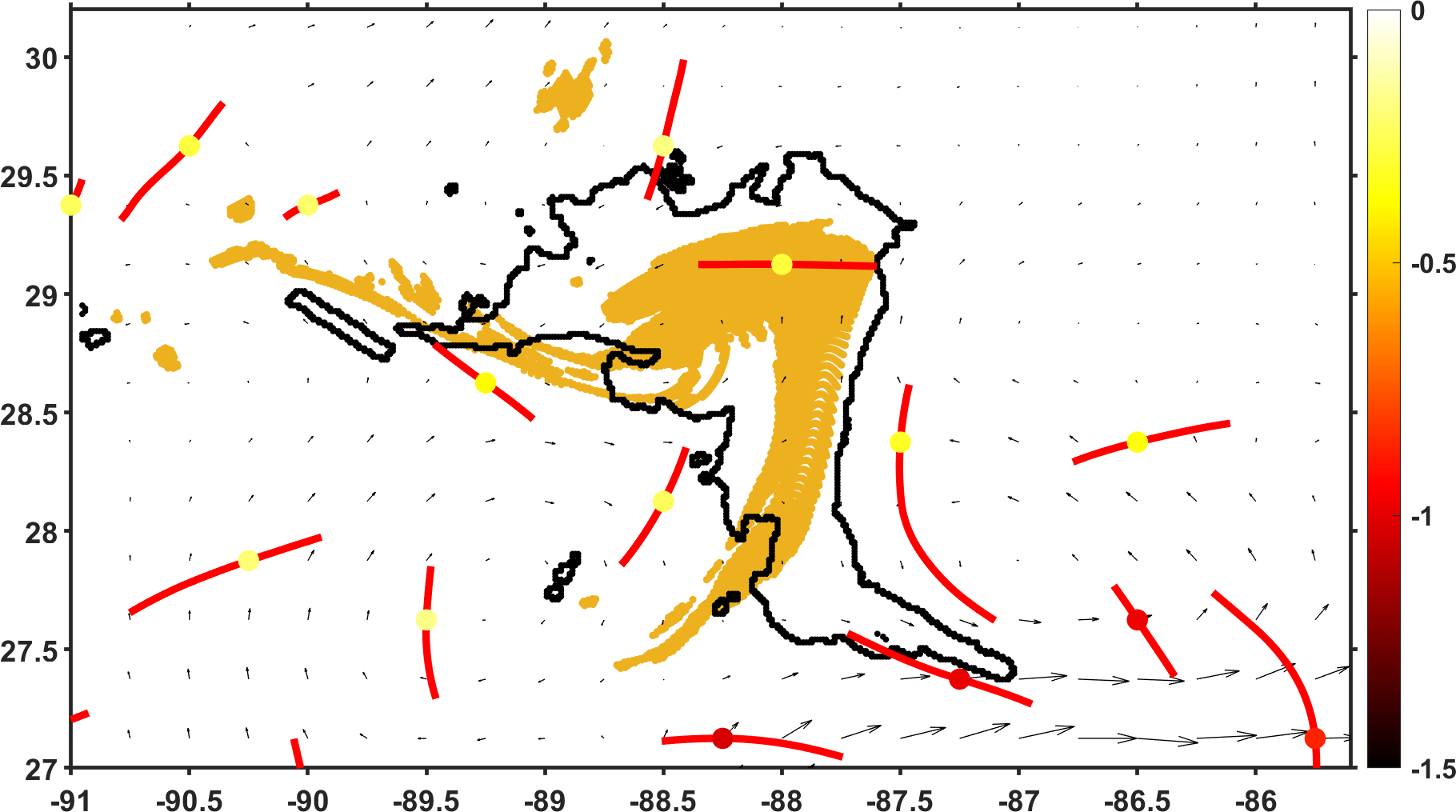}
	\caption{Same as in figure \ref{fig:trap11} but on May 15, 2010. }
	\label{fig:trap15}
	\end{center}
\end{figure}

By May 15, observed oil has aligned with one of the strongest TRAPs, the one that has remained near 27.4N and 87.25W since May 10. Meanwhile, the simulated oil trajectory continues its original path towards the southwest, by now clearly diverging from the observed path (Fig. \ref{fig:trap15}). The weak TRAP originally indicating the path towards the southwest (near 28.2N, 88.3W in Figs. \ref{fig:trap11} and \ref{fig:trap13}) is no longer present in the May 14 velocity, and will not be seen again during the rest of our analysis.

\begin{figure}[H]
	\begin{center}
\includegraphics[scale=0.37]{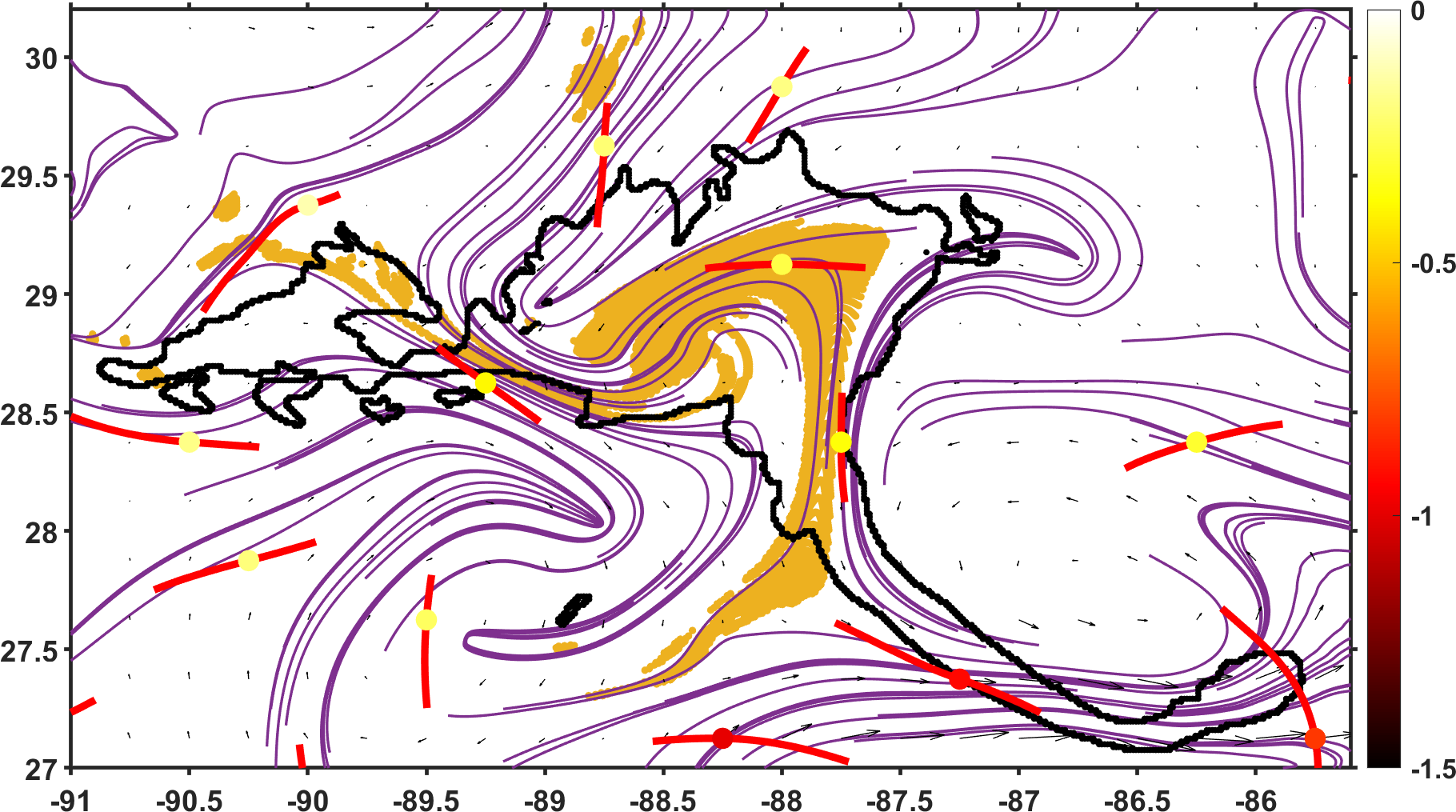}
	\caption{ Same as in figure \ref{fig:trap11} but on May 17, 2010; purple lines are attracting LCS
	computed back in time between May 17 and May 10. }
	\label{fig:trap17} %datestr(t(end-10))='14-Mar-2016 06:00:00'
	\end{center}
\end{figure}

By May 17, observed oil has deformed towards the south then east, while the simulated oil trajectory has deformed towards the south then west, thus the observed and simulated trajectories are heading in opposite directions (Fig. \ref{fig:trap17}).  LCS illustrate a hyperbolic point near 28.4N and 87.7W where transport splits, the western part heading towards the south then west (simulated oil follows this LCS) and the eastern part moving south then east (observed oil follows this LCS). Thus, LCS show that the simulated tracer just barely missed the observed transport pattern that is accurately depicted by the LCS on the eastern side of the hyperbolic point. Note there is a TRAP above the LCS on the eastern side of the hyperbolic point near 28.4N and 87.7W, accurately selecting the altimetric LCS that agrees with the path of observed oil transport. Further confirmation comes from the strong TRAP near 27.4N and 87.25W forecasting elongation of oil along the correct direction since at least May 10. This shows how TRAPs and LCS provide complementary information, together explaining the discrepancy between simulated and observed trajectories. With such a detailed Lagrangian characterization of the available velocity, an oil-spill modeler can then identify which patterns are most likely to occur and which patterns are likely spurious. In this example, a TRAP can predict the observed movement of oil at least 8 days in advance, starting from the velocity snapshot on May 10, 2010, while the simulated trajectory for oil initiated from the observed oil on May 10, 2010, is caught on the wrong side of a hyperbolic point and ends up moving in the opposite direction relative to observed transport.

Although initially the velocity is correct in inducing transport towards the southwest, the TRAP that accurately identifies southwest motion is weak and it disappears after a few days (Figs. \ref{fig:trap11}, \ref{fig:trap13} and \ref{fig:trap15}). The hyperbolic point causing the divergence of simulated transport relative to observed transport (Fig. \ref{fig:trap17}) therefore seems to originate from a disparity in the velocity arising from coarse temporal resolution and resulting in error accumulation during Lagrangian integration. The problem may be related to the period between passes of altimetry satellites being too long to capture changes in the ocean velocity on timescales of a few days, and the  attendant influence on trajectories when integrating the velocity. Fortunately, velocity from altimetry accurately captures the features that result in the main transport patterns, it is just that velocity integration is not an adequate tool to extract this information.

The above results are from the GEKCO2 velocity; very similar results are obtained using HyCOM Global (not shown). Simulated transport is also very similar when using HyCOM GoM (not shown). The computation of TRAPs for flows at high resolutions (about 4 km or less in the ocean) may require filtering and is a topic of current research. Global GEKCO2 velocity is available since 1993 to date minus two days.

%However, in this latter case, the computation of TRAPs is less reliable even after smoothing the higher resolution velocity, and using high-order differentiation for the strain-rate computation.
%\textcolor{magenta}{RODRIGO: not sure I would mention the latter sentence as we did not do an appropriate filtering. When I used the HFR velocity for Martha-s Vineyard, the filtering was done tailored to the way the velocity.}

\section{Conclusion and Outlook}
\label{sec:eight}

Recent advances suggest that better results for oil-spill modelers are a reachable goal. In this chapter, we have shown examples of how a basic understanding of the physics driving motion in the sea and the use of novel Lagrangian and Eulerian Coherent Structures techniques can result in improved oil-spill modeling.

Recent progress in Coherent Structures techniques---computing attracting structures that shape material transport from an Eulerian snapshot---is a promising development for oil-spill modeling efforts. We have shown how TRAPs (or Attracting OECSs) can bypass errors in the velocity that produce large errors in simulated trajectories. \cite{Olascoaga2012} explored similar ideas by searching for the most persistent hyperbolic cores using 15-day integrations. Lagrangian integration over such a period acts as a filter, removing short-term variability and focusing on finite-time mesoscale features.  Our results are consistent with theirs: a highly attractive hyperbolic core persists for over a week, accurately anticipating prominent fluid deformation. The advantages of TRAPs are that they do not require velocity integration, they can be computed from a single velocity snapshot, and they predict hyperbolic attraction cores whether persistent or ephemeral.

As described in section \ref{sec:example}, TRAPs were able to identify the correct transport patterns, while LCS and simulated trajectories were influenced by a deficient velocity. The erroneous simulated transport is initially correct as evidenced by the observed movement of oil,  aptly identified in the velocity by a weak, ephemeral TRAP. However, simulated trajectories become erroneous as integration causes velocity-error accumulation, while a strong persistent TRAP marked the correct region of oil confluence well in advance of observed deformation.

Higher resolution observations of sea-surface velocity and surface processes are expected to advance our understanding, ultimately resulting in improved velocity products.  For example, high-resolution observations, theoretical modeling, and coupled ocean-atmosphere-wave numerical models can be expected to improve our understanding of the ocean's surface \citep{VillasBoas2019}. Improvements in observations and understanding should translate to improved oil transport forecasts. As velocity products improve by including more of the physics relevant to simulating oil's movement, the techniques highlighted here will become more relevant. A basic understanding of ocean physics will continue to be needed to supplement velocity products lacking certain types of forcing, but also to understand new velocity products that will incorporate more types of physics than previously available. 

Despite improvements in velocity products, the sensitivity of trajectory computations to small errors will likely continue to produce erroneous trajectories. This suggests that the use of novel techniques that seek to bypass the sensitivity inherent to trajectory computations are likely to become important tools for the oil-spill modeler. Thus, as an effort that is parallel to improving velocity products, progress in techniques bypassing the problems inherent to the unstable nature of ocean currents can be expected. Examples include Objective Eulerian Coherent Structures for instantaneous transport patterns, and climatological Lagrangian Coherent Structures for climatological transport patterns. The latter is an empirical approach developed in \cite{Duran2018cLCS} where it was found that filtering the velocity by computing a climatology is surprisingly accurate for identifying recurrent Lagrangian transport patterns if the proper Lagrangian tools are used. Among their results, the transport pattern studied in section  \ref{sec:example} turns out to be a recurrent pattern, and therefore a pattern that is likely to be seen in May through August of any given year. A climatological approach should not replace forecasts, yet it does provide a valuable general understanding of persistent transport barriers, trajectories, regions of persistent attraction, and persistent isolation. Thus, Lagrangian climatologies in the sense of \cite{Duran2018cLCS} complement the interpretation of forecasts while providing a broad understanding of Lagrangian motion in a region of interest. The climatological approach suggests that progress can be made by understanding the connection between the inherently time-dependent trajectories of an instantaneous ocean velocity, and a low-pass filtered climatological velocity. An alternative approach to understanding uncertainty in oil-spill modeling is the use of ensemble simulations to create a surrogate model \citep{Zhang2020}. It is possible that future work might be able to bridge the Lagrangian climatology strategy of \cite{Duran2018cLCS} with the ensemble-based surrogate model of \cite{Zhang2020}.

As ocean observations and ocean models improve with new satellite products, an increasing number of HF radars, drifters and autonomous vehicles, and advances in data processing, assimilation, and numerical modeling,  oil-spill modelers should be able to capitalize from the material presented here, achieving a higher rate of success in forecasting and hindcasting the movement of oil.

%of interest to society
Accurately simulating Lagrangian transport in the ocean is of considerable societal interest for a variety of reasons including oil spills, the fate of other contaminants, fisheries, ocean ecology, search and rescue, tracing accidents or crimes back in time (forensic work), climate change and weather predictions, among others. Many countries have conducted oceanographic research for several decades now. Consequently, enough progress has been made to where simulating trajectories in the ocean often produces valuable information. For the needed progress to continue, we must understand the ocean's importance for society at large, and that the relevance of oceanographic endeavor is increasing due to pressing issues including climate change, coastal development, population growth, and globalization.

%An example related to weather predictions is when ocean-atmosphere coupled models try to predict hurricane intensification based on the model's sea surface temperature: did the ocean model simulate the transport of warm water correctly? Saving lives and very large sums of money depend on the answer. %For example in the Gulf of Mexico, warm tropical water advected by the Loop Current and associated eddies has a strong influence on a hurricane's intensity, and is  topic of active research.

%For example, how do you feed billions of humans if fisheries collapse? This is a very real threat we are currently facing \cite[e.g.][]{Worm2016}.

\section{Acknowledgments}

The GEKCO2 product used in this study was developed and extracted by Joël Sudre et al. at LEGOS, France. GEKCO2 data can be requested at \href{http://www.legos.obs-mip.fr/members/sudre/gekco\_form}{\nolinkurl{http://www.legos.obs-mip.fr/members/sudre/gekco\_form}}. Funding for the development of HyCOM has been provided by the National Ocean Partnership Program and the Office of Naval Research. Data assimilative products using HyCOM are funded by the U.S. Navy. Computer time was made available by the DoD High Performance Computing Modernization Program. The output is publicly available at  \href{https://hycom.org}{https://hycom.org}. Deepwater Horizon oil products from NOAA are available at
\href{http://www.ssd.noaa.gov/PS/MPS/deepwater.html}{\nolinkurl{http://www.ssd.noaa.gov/PS/MPS/deepwater.html}}. RD would like to thank M. J. Olascoaga for helpful conversations.\\

The work of RD was performed in support of the US Department of Energy’s Fossil Energy, Oil and Natural Gas Research Program. It  was executed by NETL’s Research and Innovation Center, including work performed by Leidos Research Support Team staff under the RSS contract 89243318CFE000003. M.S. acknowledges support from the Schmidt Science Fellowship and the Postdoc Mobility Fellowship from the Swiss National Foundation.

This work was funded by the Department of Energy, National Energy Technology Laboratory, an agency of  the United States Government, through a support contract with Leidos Research Support Team (LRST). Neither the United States Government  nor any agency thereof, nor any of their employees, nor LRST, nor any of their employees, makes any warranty, expressed or implied, or  assumes any legal liability or responsibility for the accuracy, completeness, or usefulness of any information, apparatus, product, or process disclosed, or represents that its use would not infringe privately owned rights. Reference herein to any specific commercial product, process, or service by trade name, trademark, manufacturer, or otherwise, does not necessarily constitute or imply its endorsement, recommendation, or favoring by the United States Government or any agency thereof. The views and opinions of authors expressed herein do not necessarily state or reflect those of the United States Government or any agency thereof.

%C://Users/Rodrigo/OneDrive/Documents/bibtex/library
%C://Users/Rodrigo/OneDrive/Documents/netl/PROPOSALS/oleg_book/Book_chapters/repo/VerticalMixingChapter/doc/references
%C://Users/Rodrigo/OneDrive/Documents/bibtex/fot,

 \bibliography{references2}

\appendix
\section{Automated oil-spill simulations}
\label{app:diff}
To simulate an oil spill with advection and diffusion but without the need to choose an eddy diffusion coefficient, the advection-diffusion equation is solved in Lagrangian terms including an automated method to determine an eddy diffusion coefficient. Mathematically, and considering two-dimensional horizontal transport, this amounts to solving a stochastic differential equation (SDE), given by
\begin{align}
    \label{eq:sde}
    \d \ve{X} = \left( \ve{u} + \nabla \kappa \right) \d t + \sqrt{2\kappa} \, \d \ve{W}(t),
\end{align}
where $\kappa$ is assumed to be a scalar function of space and time, and where the random variable $\ve{W}(t)$ is a two-dimensional Wiener process \citep[see, e.g.,][p.~28,~70]{kloeden1992}. Here, we have assumed that the diffusivity is isotropic, i.e., it is the same in both horizontal directions. For details of the anisotropic case, the interested reader is referred to \cite{spivakovskaya2007}. If we solve this equation for a sufficiently large number of particles, the density of particles will evolve in time in the same way as the concentration, $C$, described by Eq.~\eqref{eq:adv-diff} \cite[p.~122--126]{lynch2014}.

The diffusion part is typically modeled as a random walk, by numerically solving Eq.~\eqref{eq:sde}, with $\ve{u}=0$ if advection is separately accounted for. The simplest numerical scheme for SDEs is the Euler-Maruyama scheme \cite[p.~305]{kloeden1992}, which in our case (with $\ve{u}=0$) is
\begin{align}
    \label{eq:em}
    \ve{X}_{n+1} = \ve{X}_n + \left( \nabla \cdot \kappa \right) \Delta t + \sqrt{2\kappa} \, \Delta \ve{W}_n.
\end{align}
Here, $\ve{X}_n$ is the position of a particle at time $t_n$, $\Delta t$ is the time step, and $\Delta \ve{W}_n$ are the increments of the two-dimensional Wiener process. That is, $\Delta \ve{W}_n$ is a vector with two independent, identically distributed random components, with zero mean, and variance $\Delta t$. If the diffusivity is spatially variable, accounting for its gradient in Eq.~\eqref{eq:em} avoids nonphysical transport away from regions of high diffusivity \cite[p.~125]{lynch2014}. However, this problem is usually more important for vertical transport, as diffusivity gradients are usually both sharper and more persistent in the vertical \citep{Nordam2020book}.

%Gaussian random variables are a common choice, but also a uniform distribution on the interval $[-\sqrt{3 \Delta t}, \sqrt{3 \Delta t}]$ will have the correct mean and variance, and can be used.

The diffusivity can be estimated by different means, and is sometimes provided by an ocean model, but it will usually include uncertainty and errors. It is also important to remember that the eddy diffusivity does not directly correspond to any physical, measurable quantity in nature. Rather, it is a parameterization of the combined effect of unresolved eddy motion (subgrid stirring), and molecular diffusivity. Note that since the eddy diffusivity is intended to compensate for unresolved features in the ocean model, the eddy diffusivity will be higher for low-resolution models, and smaller for high-resolution models. A simple scheme suggested by \cite{smagorinsky1963general} scales the eddy diffusivity with the square of the model grid cell size, which may be a useful rule-of-thumb.

Another option is to use a time-dependent diffusivity, which increases with the
time since the release. The rationale for this approach is found in
observations.  In the ocean, eddies exist at a wide range of spatial scales,
from the largest basin-scale gyres, down to the Kolmogorov length scale of
millimeters or less.  The effect of these eddies on a patch of dissolved
tracer depends on the size of the eddy, relative to the size of the patch.
Eddies that are much larger than the patch will only advect it, with little or
no change to its shape. Eddies that are much smaller than the patch will only
serve to deform its surface, without changing its overall shape. Eddies that
are of the same size as the patch, on the other hand, will significantly change
its shape, by stretching out filaments in different directions, thus increasing
the overall size of the patch.

A small patch will initially be most affected by small eddies, but as it grows in
size, it will be affected by increasingly large eddies. By constructing an
argument based on the typical turnover time of eddies of different sizes, it is
possible to arrive at an expression for how fast the size of the patch will grow
in time. Following the argument of \cite[pp.~257--258]{davidson2015}, we let
$R$ be the mean radius of an initially small and spherical patch, and let the
typical speed of an eddy of size $r$ be $v_r \sim (\varepsilon r)^{1/3}$, where
$\varepsilon$ is the turbulent kinetic energy dissipation rate per unit mass.
Since the patch is mainly affected by eddies of its own size, we get
\begin{align}
    \label{eq:dRdt}
    \frac{\ud R}{\ud t} \sim v_R \sim (\varepsilon R)^{1/3}.
\end{align}
Rewriting this expression by using that $\frac{\ud}{\ud t}R^2 = 2R \frac{\ud
R}{\ud t}$, we get
\begin{align}
    \label{eq:fourthirds}
    \frac{\ud R^2}{\ud t} \sim \varepsilon^{1/3} R^{4/3},
\end{align}
which is known as Richardson's four-thirds law \cite{richardson1926}. This
expression is only valid for $\eta \ll R \ll \ell$, where $\eta$ is the
scale of the smallest eddies (the Kolmogorov scale), and $\ell$ is the
scale of the largest eddies \cite[p.~258]{davidson2015}. A further limitation
in our case is that on large scales, the ocean is essentially two-dimensional.
We will return to this point.

Since $R^2$ is proportional to the variance of a patch of tracer, we see that the
rate of increase of the variance is size-dependent, and thus time-dependent,
when a patch is subject to turbulent mixing. This is contrary to the case in
Fickian diffusion described by Eq.~\eqref{eq:diff}, where the variance grows
linearly with time, proportional to the diffusivity:
\begin{align}
    \label{eq:variance}
    \frac{\ud R^2}{\ud t} \sim \kappa.
\end{align}
From the above, we can derive a time-dependent ``effective diffusivity'',
$\kappa_{\textrm{eff}}(t)$, for a patch subject to turbulent mixing. Integrating
Eq.~\eqref{eq:dRdt}, we find that $r \sim \varepsilon^{1/2} t^{3/2}$, and
inserting this into Eq.~\eqref{eq:fourthirds} we find
\begin{align}
    \label{eq:Keff}
    \kappa_{\textrm{eff}}(t) \sim \varepsilon t^2.
\end{align}
Hence, we find that the variance of a patch subject to turbulent mixing is
proportional to $t^3$, since it grows at a rate proportional to $t^2$.

Early experimental investigation of the above results include observations of
balloons released into the atmosphere \cite{richardson1926}, and bits of
parsnip thrown into a loch by Richardson's cabin in Scotland \citep{richardson1948}.
\cite{okubo1971} published a summary of several dye release experiments,
covering spatial scales from \SI{100}{\meter} to \SI{10}{\kilo\meter} and time
scales from hours to several weeks. When plotting variance as a function of time
(Fig.~1 in \cite{okubo1971}), he found $R^2 \sim t^{2.3}$, and when plotting
effective diffusivity as a function of spatial scale, he found $\kappa_{\textrm{eff}}
\sim R^{1.1}$. These results were later expanded with more observational data,
by, \emph{e.g.}, \citet{murthy1976} and \citet{lawrence1995}, still showing
approximately the same trends.

If we for the moment accept sloppy notation with respect to units, an explicit
expression for the time-dependent apparent horizontal diffusivity, $\kappa_a$, may
be obtained from Eq. (3) in \cite{okubo1971},
\begin{align}
    \label{eq:sigmarc}
    \sigma_{rc}^2 = 0.0108 \cdot t^{2.34},
\end{align}
where $\sigma_{rc}^2$ is measured in \si{\centi\meter\square} and $t$
is in seconds. Combining this with the relation $\kappa_a = \sigma_{rc} / 4t$, we get
\begin{align}
    \label{eq:Ka}
    \kappa_a = 0.0027 \cdot t^{1.34},
\end{align}
where $\kappa_a$ is given in units \si[per-mode=symbol]{\centi\meter\square\per\second}. These observation-based results do not agree with the theoretical considerations summarised in Eq.~\eqref{eq:fourthirds}. However, it is clear that the ocean cannot be considered to be three-dimensional when considering a
patch of size \SI{100}{\meter} or above, released in the mixed layer. Hence, the theoretical results cannot be expected to hold exactly.

Based on the discussion above, it might seem reasonable to use a time-dependent
diffusivity in an oil spill model. However, it is important to remember that the
effective diffusivity is intended to mimic the mixing due to eddies in the ocean
currents. If high-resolution current data is used as input to the modeling,
more of those eddies will already be represented in the data, and need not be
accounted for in the time-dependent diffusivity. Hence, the diffusivity should
in some sense be matched to the resolution of the ocean current data.

If the horizontal resolution of the current data is $\Delta x$, then any patch
of tracer with $R \ll \Delta x$ will only be advected along the currents,
without changing its shape significantly. Hence, it makes sense to apply a
time-dependent diffusivity to small patches. However, once the patch grows in
size such that $R > \Delta x$, differential advection by eddies represented in
the current data will lead the patch to spread out further. Applying an
additional time-dependent diffusivity to such a patch will then lead to too much
diffusion.

In practice, it is easier to truncate the effective diffusivity based on
time, rather than spatial scales. It is a simple matter to keep track
of the ``age'' of numerical particles, and use a time-dependent diffusivity in
the random walk for each particle. Future work will be need to determine when to truncate the time-dependent diffusivity, and to quantify the difference of time-dependent diffusivity instead of a constant one in practical applications. Further reading can be found in \citet[][chapter IV]{csanady1973}, \citet[][chapter 2]{okubo2013}, and \citet[][chapter 4]{lynch2014}.

\end{document}